\documentclass[12pt,preprint,dvips,deluxetable]{aastex}
\def \msun{$\mathrm{M}_\odot$}
\def \rsun{$\mathrm{R}_\odot$}

\def \kms{km~s$^{-1}$}
\def \1pap{Paper I}
\def \pap2{Paper II}

\makeatletter
\newcommand\cutinheadb[1]{%
 \noalign{\vskip .8ex}%
 \@ptabularcr
 \noalign{\vskip -4ex}%
 \multicolumn{\pt@ncol}{c}{#1}%
 \@ptabularcr
 \noalign{\vskip .8ex}%
 \hline
 \@ptabularcr
 \noalign{\vskip -1.5ex}%
}%
\def\cutinheadb@ppt#1{%
 \noalign{\vskip .8ex}%
 \@ptabularcr
 \noalign{\vskip -1.5ex}% Style Note: in apj, it is -1.5ex
 \multicolumn{\pt@ncol}{c}{#1}%
 \@ptabularcr
 \noalign{\vskip .8ex}%
 \hline
 \@ptabularcr
 \noalign{\vskip -1.5ex}%
}%
\makeatother

\author{Daniel~C. Kiminki\altaffilmark{1},
Henry~A. Kobulnicky\altaffilmark{1}, Ian Gilbert\altaffilmark{2}, Sarah
Bird\altaffilmark{3}\altaffilmark{4}, Georgi Chunev\altaffilmark{5}}

\altaffiltext{1}{Dept. of Physics \& Astronomy, University of Wyoming,
Laramie, WY 82070} 
\altaffiltext{2}{Dept. of Physics, Grove City College, Grove City, PA 16127}
\altaffiltext{3}{Dept. of Physics \& Astronomy, University~of~Turku, 
Turku, Finnland}
\altaffiltext{4}{Tuorla Observatory, Piikki\"{o}, Finnland}
\altaffiltext{5}{Dept. of Computer Science, Indiana University, Bloomington, IN 47405}

\begin{document}

\title{Five More Massive Binaries in the Cygnus OB2 Association}

\begin{abstract}
We present the orbital solutions for four OB spectroscopic binaries,
MT145, GSC 03161-00815, 2MASS J20294666+4105083, and Schulte~73, and
the partial orbital solution to the B spectroscopic binary, MT372, as
part of an ongoing study to determine the distribution of orbital
parameters for massive binaries in the Cygnus OB2 Association. MT145
is a new, single-lined, moderately eccentric ($e=0.291\pm0.009$)
spectroscopic binary with period of $25.140\pm0.008$~days. GSC
03161-00815 is a slightly eccentric ($e=0.10\pm0.01$), eclipsing,
interacting and double-lined spectroscopic binary with a period of
$4.674\pm0.004$~days. 2MASS J20294666+4105083 is a moderately
eccentric ($e=0.273\pm0.002$) double-lined spectroscopic binary with a
period of $2.884\pm0.001$~days. Schulte~73 is a slightly eccentric
($e=0.169\pm0.009$), double-lined spectroscopic binary with a period
of $17.28\pm0.03$~days and the first ``twin'' in our survey with a
mass ratio of $q=0.99\pm0.02$. MT372 is a single-lined, eclipsing
system with a period of $2.228$~days and low eccentricity
($e\sim0$). Of the now 18 known OB binaries in Cyg OB2, 14 have
periods and mass ratios. Emerging evidence also shows that the
distribution of $log(P)$ is flat and consistent with ``\"{O}pik's Law''.
\end{abstract}

\keywords{ techniques: radial velocities --- (stars:) binaries:
 general --- (stars:) binaries: spectroscopic --- (stars:) binaries:
 (\textit{including multiple}) close --- stars: early-type --- stars:
 kinematics --- surveys }

\section{Introduction}
The study of massive binary systems in young clusters plays a key role
in our understanding of how massive stars form. In addition to
providing the binary fraction for massive stars, massive binary
systems provide information about their formation environment in the
form of quasi-preserved parameters such as eccentricity, separation,
period, and angular momentum \citep[in the form of rotational
velocities;][]{Larson01}. In addition, an initial mass function (IMF)
composed of the secondary component masses can indicate whether the
companions are randomly drawn from a \citet{Salpeter55} (or equivalent
field star) IMF or not. This can tell us whether the binary systems may
have randomly formed by way of gravitational capture after the
formation process, or preferentially paired through a more
complicated formation process such as competitive accretion
\citep{Bonnell98}.

Cyg OB2 provides one of the best regions for indirectly examining the
formation of massive stars as it houses $\sim$60--70 O-type stars
\citep{Neg08}, including an O3If \citep[Schulte~7;][]{Walborn00} and
an O4If \citep[Schulte~22;][]{Comeron07}, and possibly more than 2000
B stars \citep{Kn00}. \citet{Kiminki08} presented six new
spectroscopic binary systems and summarized the then-known 11 OB
binary systems in Cyg OB2. Between then and this work, three
additional OB binary systems have been uncovered, including Schulte~9
\citep{Naze08}, 2MASS J20302730+4113253 \citep[][submitted]{Stroud09},
and GSC~03161-00815 \citep{NSVSa,Hanson03}. Six of the total 14
systems, MT421, MT429, MT696, Schulte~3, Schulte~5, and
GSC~03161-00815 are identified as eclipsing binaries (where notation
for the first five are from \citealt{MT91} and
\citealt{Schulte56}). Two of the 14 systems, MT059 and MT258, are
single-lined spectroscopic binaries (SB1s), and eight of the 14,
MT252, MT696, MT720, MT771, Schulte~3, Schulte~8a, Schulte~9, and
2MASS J20302730+4113253, are double-lined spectroscopic binaries
(SB2s). In part three of this ongoing study, we add to the 14 and
present the newly uncovered spectroscopic binaries, MT145 (SB1),
2MASS~J20294666+4105083 \citep[listed as a candidate SB2
in][]{Hanson03}, Schulte~73 (SB2), and MT372 (SB1). We also present
the first spectroscopic solution to the eclipsing system,
GSC~03161-00815 (SB2). For simplicity, we will use \citet{Comeron02}
notation for GSC~03161-00815, 2MASS~J20294666+4105083, and 2MASS
J20302730+4113253 (A36, A45, and B17 respectively). These new systems
bring the total number of OB binaries in Cyg OB2 to 18, constituting
one of the highest numbers of massive binary solutions of any open
cluster.

Section~2 of this work provides observational details of the new
spectroscopic datasets. Section~3 discusses the measurement of radial
velocities, the search for periods in the radial velocity data, and
the determination of orbital elements via radial velocity curve
fitting. Section~4 discusses the orbital solutions to the SB1, MT145,
and the SB2s, A36, A45, and Schulte~73 (Cyg OB2 No.~73). Section~5
presents the partial solution to the SB1, MT372. Finally, Section~6
summarizes the results of the survey to date, including the total
number of OB binaries uncovered in the Cyg~OB2 core region, the total
number of O~star binary solutions, and the emerging distribution of
orbital periods.

\section{Observations}
\citet{Kiminki07} \&\ \citet{Kiminki08} (Papers I \&\ II) detail
the observations of this survey through 2007 September. We have
obtained additional datasets with the WIRO-Longslit
spectrograph\footnote{The WIRO-Longslit spectrograph was constructed
by M. J. Pierce and A. Monson at the University of Wyoming. Instrument
information can be found at
http://physics.uwyo.edu/$\sim$amonson/wiro/long\_slit.html} on the
Wyoming Infrared Observatory (WIRO) 2.3~m telescope and the Hydra
spectrograph on the WIYN\footnote{The WIYN Observatory is a joint
facility of the University of Wisconsin, Indiana University, Yale
University, and the National Optical Astronomy Observatory} 3.5 m
telescope. Table~\ref{obs.tab} lists the observing runs at each
facility, the corresponding spectral coverages, and mean spectral
resolutions.

Observations at WIYN took place over six nights on 2008 June
10--15. We used the Hydra spectrograph with the Red camera, 2\arcsec\
blue fibers, and the 1200 l mm$^{-1}$ grating in second order to
obtain four 1500~s exposures in each of two fiber configurations
yielding a maximum signal-to-noise ratio (SNR) of 80:1 for the
brightest stars. The spectral coverage was 3820--4500~\AA\ at a mean
resolution of $R\sim$4500. Copper-Argon lamps were used between each
exposure to calibrate the spectra to an RMS of 0.03~\AA\ (2 \kms\ at
4500~\AA), and the typical resolution was 1.0~\AA\ FWHM at 3900~\AA\
and 0.82~\AA\ FWHM at 4400~\AA. Spectra were Doppler corrected to the
heliocentric frame and checked against the radial velocity standards
HD131156 (G8V), HD146233(G2V), HD161096(K2III), HD161797(G5IV), and
HD171391(G8III) from \citet{Stefanik1999} before comparison to
previous datasets.

Observations using the WIRO-Longslit spectrograph with the
1800~l~mm$^{-1}$ grating in first order took place over 37 nights
between 2007 October 23 and 2008 September 19 to examine the
H$\alpha$, \ion{He}{1}, and \ion{He}{2} absorption lines in suspected
SB2s. Exposure times varied from 600~s to 4500~s (in multiples of
600--900~s) depending on weather conditions and yielded a maximum SNR
of 200:1 for the brightest stars. The spectral coverages were
5500--6750~\AA\ (2007 October 23 through 2007 November 5) and
5250--6750~\AA\ (2008 June 23 through 2008 September 19). Copper-Argon
lamp exposures were taken after each star exposure to wavelength
calibrate the spectra to an rms of 0.03~\AA\ (1.4~\kms\ at
6400~\AA). The typical spectral resolution was 1.5~\AA\ FWHM across
the chip. Spectra were Doppler corrected to the heliocentric frame and
checked against the same radial velocity standards taken during the
2008 June observations at WIYN before comparison to previous
datasets. In addition, we also cross-correlated the WIRO-Longslit
spectra with a composite interstellar line spectrum to look for
systematic relative radial velocity shifts. The interstellar line
spectrum was created by extracting the interstellar lines present in
the 2008 spectra of A36, Doppler shifting them to the same velocity
and combining them. Interstellar lines and diffuse interstellar bands
were identified with \citet{Herbig1975}, \citet{Herbig1991},
\citet{Morton1991}, and \citet{Gala00}. Relative velocity shifts were
generally less than 6~\kms. Observations with relative shifts larger
than 6~\kms\ were corrected to the average absolute radial velocity
offset before computing orbital solutions.  All new datasets were
reduced using standard IRAF reduction routines as outlined in Paper I.

\section{Data Analysis and Orbital Solutions}
\subsection{Measuring Radial Velocities and Estimating Errors}
We obtained radial velocities, $V_r$, for the single-lined system,
MT145 via the IRAF cross-correlation task XCSAO in the RVSAO package
\citep{xcsao}, using a model stellar atmosphere
\citep[TLUSTY]{LHub2003} of the appropriate effective temperature and
gravity as discussed in Paper I. Errors within the XCSAO
task are calculated using,

\begin{equation}
\sigma_v = \frac{3w}{8(1+r)}.
\end{equation}

\noindent where $w$ is the FWHM of the correlation peak and
$r$ is the ratio of the correlation peak height to the amplitude of
antisymmetric noise \citep{xcsao}. 

To obtain rough $V_r$ measurements for the double-lined systems, A36,
A45, and Schulte~73, we deblended the \ion{He}{1}
$\lambda\lambda$4471, 5876 (A36 \&\ Schulte~73) or H$\alpha$ (A45 \&\
MT372) lines by fitting simultaneous Gaussian profiles (at fixed
Gaussian widths) with the SPLOT routine in IRAF. We repeated this
method 10 times while varying the baseline region used to define the
continuum each time. We then used the mean of each component's list of
Gaussian centers to compute the velocity. Utilizing this method, a
number of factors contribute to the radial velocity uncertainty. These
factors include, but are not limited to the errors associated with
Gaussian profile fitting, wavelength calibrations, emission present in
the line cores, and lower signal-to-noise of the less-luminous
component's spectral profiles (especially while in a partially or
wholly blended state). We estimated the contribution of the Gaussian
profile fitting uncertainty by adopting the rms of each component's
list of Gaussian centers.  Unfortunately, this only produces a lower
limit on the uncertainty because the initial Gaussian width estimates
also affect the center measurements. For the total one-sigma
uncertainties, we adopted these rms values combined in quadrature with
the typical error in the wavelength calibration (1.4~\kms\ for WIRO
and 2~\kms\ for WIYN) and the typical rms observed in the list of
velocities obtained from the correlation with the interstellar line
template.  The total error given for each system reflects a lower
limit as we cannot characterize uncertainty contributions from
emission present in the line cores (e.g., A36) or the lower
SNR of the secondary spectral features (e.g., A45).

\subsection{Determining Orbital Solutions}
We used the method outlined in \citet{mcswain03} to determine the
orbital parameters of each system. We obtained an estimate of the
orbital period through an IDL\footnote{The Interactive Data Language
(IDL) software is provided by ITT Visual Information Solutions.}
program written by A. W. Fullerton which makes use of the discrete
Fourier transform and CLEAN deconvolution algorithm of
\citet{Roberts87}. The strongest peaks in the power spectrum of each
star were examined by folding the data at the corresponding period and
inspecting the $V_r$ curve visually. Orbital elements were then
procured by using the best period as an initial estimate in the
nonlinear, least-squares curve fitting program of \citet{Morbey74}.
The best solutions were attained by manually varying the initial
guesses of key orbital parameters until a minimum in the $rms$ of the
fit was found. Weights for each point were assigned as the inverse of
the 1$\sigma$ $V_r$ error.

While it is not uncommon to quote orbital period errors to one ten
thousandth of a day (e.g.,
\citealt{Williams08,Penny08,Linder07,mcswain07,Hillwig06}), we
performed two additional tests to evaluate the appropriateness of the
orbital period uncertainties for each system. In the first test, we
explored the confidence range for each orbital parameter by fitting
the radial velocity data 1000 times, with the orbital parameters as
free variables, while adding Gaussian noise to the data. The noise
added to each data point was characterized by the one-sigma
uncertainty. We compared the resultant rms deviations to the
single-fit error estimates for each orbital parameter. The result of
this test suggested that the majority of the single-fit parameter
uncertainties were reasonable, as the Monte Carlo test routinely
produced similar uncertainties. The disadvantage to this test,
however, is that we could not examine each fit to see if it was
reasonable (i.e., whether the newly folded radial velocities had a
reasonable rms). Therefore, with the error in the period strictly in
mind, we ran the orbital fitting procedure 50--100 times while
incrementing the period by the single-fit error estimate and monitored
how the rms changed. The result of this test suggests that we, at
most, underestimate the orbital period errors by a factor of 10. With
this in mind, the errors for the orbital period throughout this work
have been increased by a factor of ten.

The complete list of orbital elements for each binary system appears
in Table~\ref{orbparms.tab}. Listed within the table are the period in
days (\textit{P}), eccentricity of the orbit (\textit{e}), longitude
of periastron in degrees (\textit{$\omega$}), systemic radial velocity
(\textit{$\gamma$}), epoch of periastron (\textit{$T_0$}), primary and
secondary semi-amplitudes (\textit{$K_1$} \&\ \textit{$K_2$}), adopted
or calculated minimum primary and secondary masses in solar masses
(\textit{$M_1$} \&\ \textit{$M_2$}), primary and secondary mass
functions in solar masses (\textit{f(m)$_1$} \&\ \textit{f(m)$_2$}),
spectral classifications from this survey (\textit{S.C.$_1$} \&\
\textit{S.C.$_2$}), the minimum primary and secondary semi-major axes
in solar radii ($a_1$sin~$i$ \&\ $a_2$sin~$i$), and finally, the rms
of the fits (\textit{rms$_1$} \&\ \textit{rms$_2$}).

\subsection{A Note on Systemic Velocities}
Systemic radial velocities are known to vary depending on which lines
are used to measure the radial velocity. This is especially evident
with massive stars and evolved stars where the stellar atmosphere
velocity gradient may be higher (see \citealt{Linder07} for examples
with a few well-studied O binaries). To minimize such discrepancies,
radial velocities are normally averaged over many lines. In the case
of MT145, where we obtain radial velocities by cross correlation over
many lines, this is not an issue. However, in the case of the
remaining 4 systems, we only have 1--2 lines (\ion{He}{1}
$\lambda\lambda$ 5876, 6678 or H$\alpha$) available for radial
velocity measurement in the majority of spectra. This places an added
uncertainty on the systemic velocity and semi-amplitudes measured for
A36, A45, Schulte~73, and MT372. In the simplest case, these effects
may cause a systematic blueshift to the systemic velocity (such as the
case with a non-eclipsing system with equal and symmetric stellar
winds). Eclipses, uneven stellar winds, and uneven surface
temperatures complicate the matter, and these effects can affect the
components unequally, producing systemic radial velocities that vary
from primary to secondary. We see this effect in two of our systems,
A45 and Schulte~73.  The component orbital solutions produce two
different systemic velocities. For both of these systems, we adopted
the mean of the two values as the systemic velocity of the system for
Table~\ref{orbparms.tab}.

\section{Orbital Solutions for MT145, A36, A45, \&\ Schulte~73}
\subsection{The SB1 MT145}
\begin{figure}
\epsscale{1.0}
\plotone{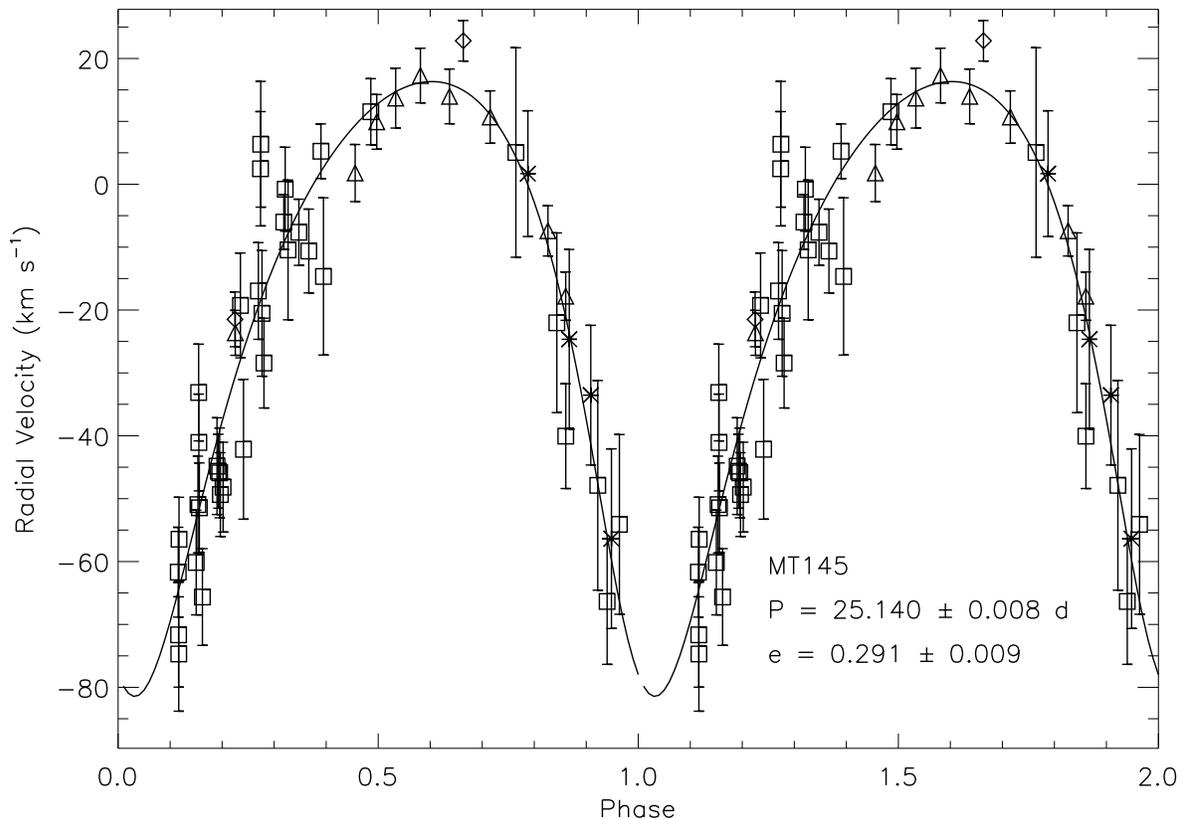}

\caption{$V_r$ curve and orbital solution to the O9III star, MT145. The
symbols correspond to the different observatories, where the asterisks are
observations taken with the Hamilton Spectrograph (Lick), the diamonds
are observations taken with HIRES (Keck), the triangles are
observations taken with Hydra (WIYN), and the squares are observations
taken with WIROspec or the WIRO-Longslit Spectrograph.
\label{MT145curve}}
\end{figure}

We observed MT145 a total of 54 times, including 6 at Lick, 2 at Keck,
12 at WIYN, and 34 at WIRO. Excluding two low-SNR spectra from Lick
and one from WIRO, a total of 51 spectra were of sufficient SNR to
obtain the solution for this O9III star. The strongest signal in the
CLEANed power spectrum corresponds to a period of 25.12 days.  An
alias at 20 days is also present. However, the 20 day period produced
a folded $V_r$ curve having a much larger rms than the 25.12~day
period. The best-fitting orbital solution and $V_r$ curve for MT145 is
shown in Figure~\ref{MT145curve}, corresponding to a period of
$P=25.140\pm0.008$ days with an eccentricity of $e=0.291\pm0.009$ and
solution $rms=8.9$~\kms. With this period, a semi-amplitude of
$K_{1}=48.9\pm0.7$~\kms, and an assumed primary mass of
$M_{1}=22.6\pm0.5$~\msun\ \citep{FM05}, we estimate a minimum
secondary mass of $M_{2}=6.0\pm0.1$~\msun, corresponding to a mid B
star \citep[interpolated from][]{drilling}. This yields a mass ratio
lower limit of $q=M_2/M_1\gtrsim0.27\pm0.01$. To constrain the mass
ratio upper limit, we examined the spectra taken while the system was
near quadrature for evidence of line-doubling or asymmetries
originating from the secondary. Neither is present in any of the
spectra. Given SNR ratios of up to 150:1 for these spectra, we would
expect to see evidence of the secondary star's spectral features at
velocity separations of $(1+{1/q})K_{1}$ if the component luminosity
ratio is larger than $L_{2}/L_{1}\simeq0.2$. This maximum luminosity
ratio is determined by combining synthetic spectra at various
luminosity ratios and line separations. The absence of secondary
spectral features therefore conservatively limits the luminosity ratio
to $L_{2}/L_{1}\lesssim0.2$ and implies a secondary less luminous than
a B1V \citep[interpolated from][]{drilling}. We thus constrain the
mass of the secondary to $6.0\pm0.1$~\msun$< M_2 \lesssim 13.8\pm
0.6$~\msun\ and the mass ratio to $0.26\lesssim q \lesssim 0.63$
(where the range in mass ratio incorporates both observational and
inclination uncertainties). These limits loosely require $i\gtrsim
31^{\circ}$ given the uncertainty in O star masses
\citep{Massey05}. Additionally, the solution produces a minimum
primary semi-major axis of $a_{1}$sin$i=23.2\pm0.3$~\rsun\ and a
longitude of periastron of $\omega=158.3^{\circ}\pm0.6^{\circ}$

The ephemerides for MT145 and MT372 are listed in Table~\ref{OC} and
include the date (\textit{$HJD-2,400,000$}), phase (\textit{$\phi$}),
measured radial velocity (\textit{$V_r$ }), cross-correlation
$1\sigma$ error (\textit{$1\sigma$~err}), and the observed minus
calculated velocity (\textit{$O-C$}).

\subsection{The SB2 A36}
A36 is originally introduced as a probable massive member of Cyg OB2
in \citet{Comeron02} and listed as a possible binary based on the
asymmetry in the absorption lines of a high SNR spectrum in
\citet[][Figure~1]{Hanson03}. \citet{NSVSa} provided a partial
photometric solution to this system and listed it as an eclipsing
binary of the Algol type with evidence of an O'Connell effect \citep[a
light curve having unequal out-of-eclipse maxima;][]{Milone1968}. We
provide the first spectroscopic solution to this double-lined system.

\begin{figure}
\epsscale{1.0}
\plotone{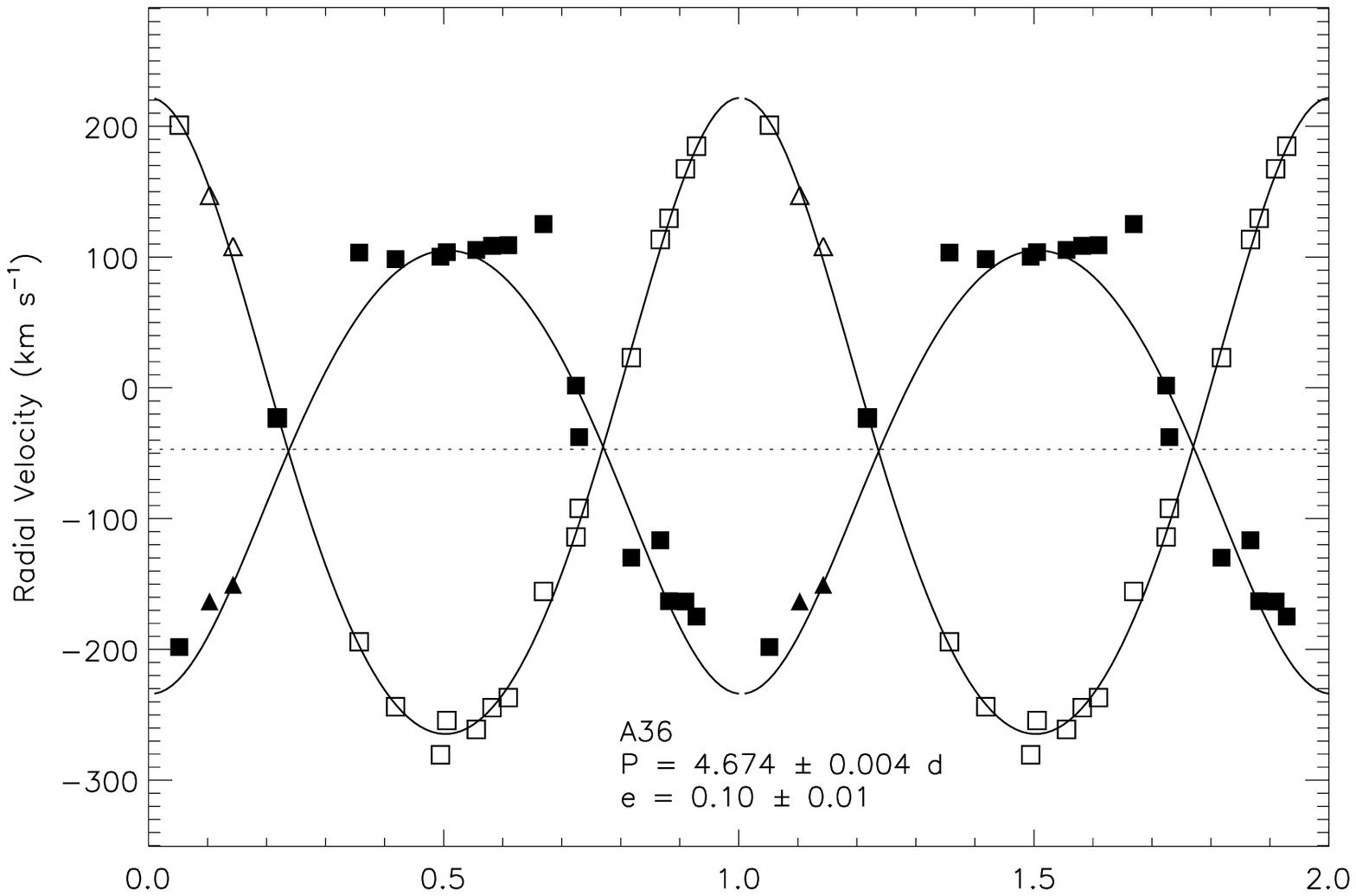}
\caption{$V_r$ curve and orbital solution for A36 using 19 of the
23 observations. The filled points correspond to the primary (B0Ib) and
the open points correspond to the secondary (B0III). The symbols represent
the same observation locations as in Figure~1.
\label{A36curve}}
\end{figure}

We observed A36 a total of 23 times between 2007 October and 2008
August, including 5 times with Hydra at WIYN and 18 times with the
WIRO-Longslit spectrograph. With the 19 higher SNR spectra we obtained
a period estimate of 4.680~days from the CLEANed power spectrum with
no aliases. From this estimate we obtained the solution for the
secondary (rms=10.0~\kms) shown in Figure~\ref{A36curve} with a period
of $P=4.674\pm0.004$ and relatively low eccentricity of
$e=0.10\pm0.01$. The period, eccentricity, epoch of periastron, and
systemic velocity were applied as fixed parameters to obtain the
solution for the primary (rms=29.3~\kms). The filled symbols
correspond to the primary $V_r$ measurements and the open symbols
correspond to the secondary $V_r$ measurements. The combined solutions
provide minimum semi-amplitudes of $K_1=169\pm9$ \kms\ and
$K_2=243\pm2$ \kms\ for the primary and secondary, respectively, and
produce semi-major axes of $a_1$sin~$i=15.6\pm0.8$~\rsun\ and
$a_2$sin~$i=22.345\pm0.003$~\rsun. The computed lower limit masses of
$M_1$sin$^3i=19.8\pm0.9$~\msun\ and $M_2$sin$^3i=13.8\pm0.9$~\msun\
yield a mass ratio of $q=0.70\pm0.06$. Table~\ref{OC2} provides the
ephemerides for A36, A45, and Schulte~73, listing the date
(HJD$-$2,400,000), phase ($\phi$), measured radial velocities
($V_{r1}$ and $V_{r2}$), and observed minus calculated velocities
($O_{1}-C_{1}$ and $O_{2}-C_{2}$). Error estimates are shown in
parentheses.

Based on two of the spectra obtained during the 2008 WIYN run, we
tentatively classify the components of A36 as a B0Ib and B0III for
primary and secondary respectively. The exposures obtained on 2008
June 10 \&\ 15 show the components at nearly the same phase and in
quadrature. Additionally, the light curve shown in \citet{NSVSa},
tells us that the temperatures of the components are similar, and the
double-lined nature of the spectra tell us that the luminosities are
similar. We can therefore conclude from the 2008 June 10 spectrum
shown first from the top in Figure~\ref{threeplot} that the spectral
types of the components are similar with subtle differences, such as
stronger \ion{Si}{4}~$\lambda\lambda$~4089,~4116 and weaker
\ion{He}{1} $\lambda\lambda$~4026,~4121,~4387,~4471 for the
primary. Both primary and secondary have no \ion{Mg}{2}~$\lambda$~4481
\citep[confirmed in Figure~1 in ][]{Hanson03}, moderately strong
\ion{Si}{4}~$\lambda\lambda$~4089,~4116, strong
\ion{He}{1}~$\lambda\lambda$~4026,~4471, weak
\ion{He}{1}~$\lambda\lambda$~4121,~4387, and strong
\ion{C}{3}~$\lambda$~4070. After smoothing the WIYN spectra to
approximately the resolution of the spectra in the \citet{WF90}
digital atlas, we combined a B0Ib (HD164402) spectrum with a B0III
(HD48434) spectrum at the appropriate radial velocities to make a
comparison. The resulting combined spectrum (second from top) is also
shown in Figure~\ref{threeplot} along with the 2001 September 8
spectrum of Schulte~73 (second from bottom) obtained at WIYN and the
associated composite spectrum for Schulte~73 (bottom). Intrinsic
\ion{He}{1}, \ion{He}{2}, \ion{C}{3}, \ion{Si}{4}, \ion{Mg}{2}, and
hydrogen absorption are labeled. The composite spectrum is not an
exact match to A36 (note that \ion{Si}{4}~4089 is stronger than that
of the primary). However, the mass ratio obtained from
\citet{drilling} for a B0Ib primary and B0III secondary is $q=0.8$ and
agrees with our calculated mass ratio of $q=0.70\pm0.06$.

\begin{figure}
\epsscale{1.0}
\plotone{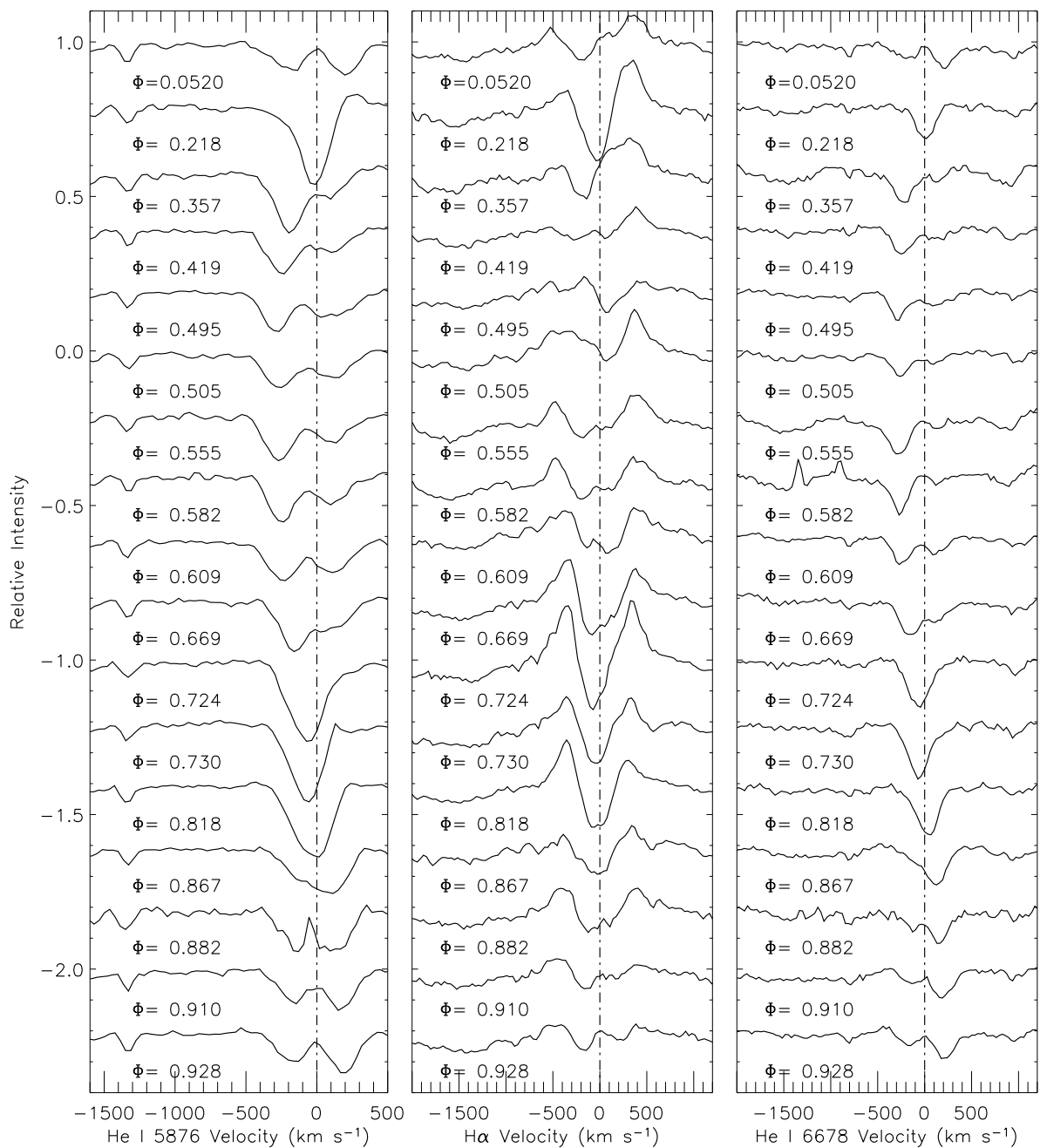}
\caption{\ion{He}{1}~$\lambda$5875.75 (left),
H$\alpha$~$\lambda$6562.80 (middle),
\ion{He}{1}~$\lambda$6678.15 (right) in velocity space and in order of phase for
observations of A36 taken at WIRO between 2007 October and 2008 August.
\label{A36progress}}
\end{figure}

\begin{figure}
\epsscale{0.9}
\plotone{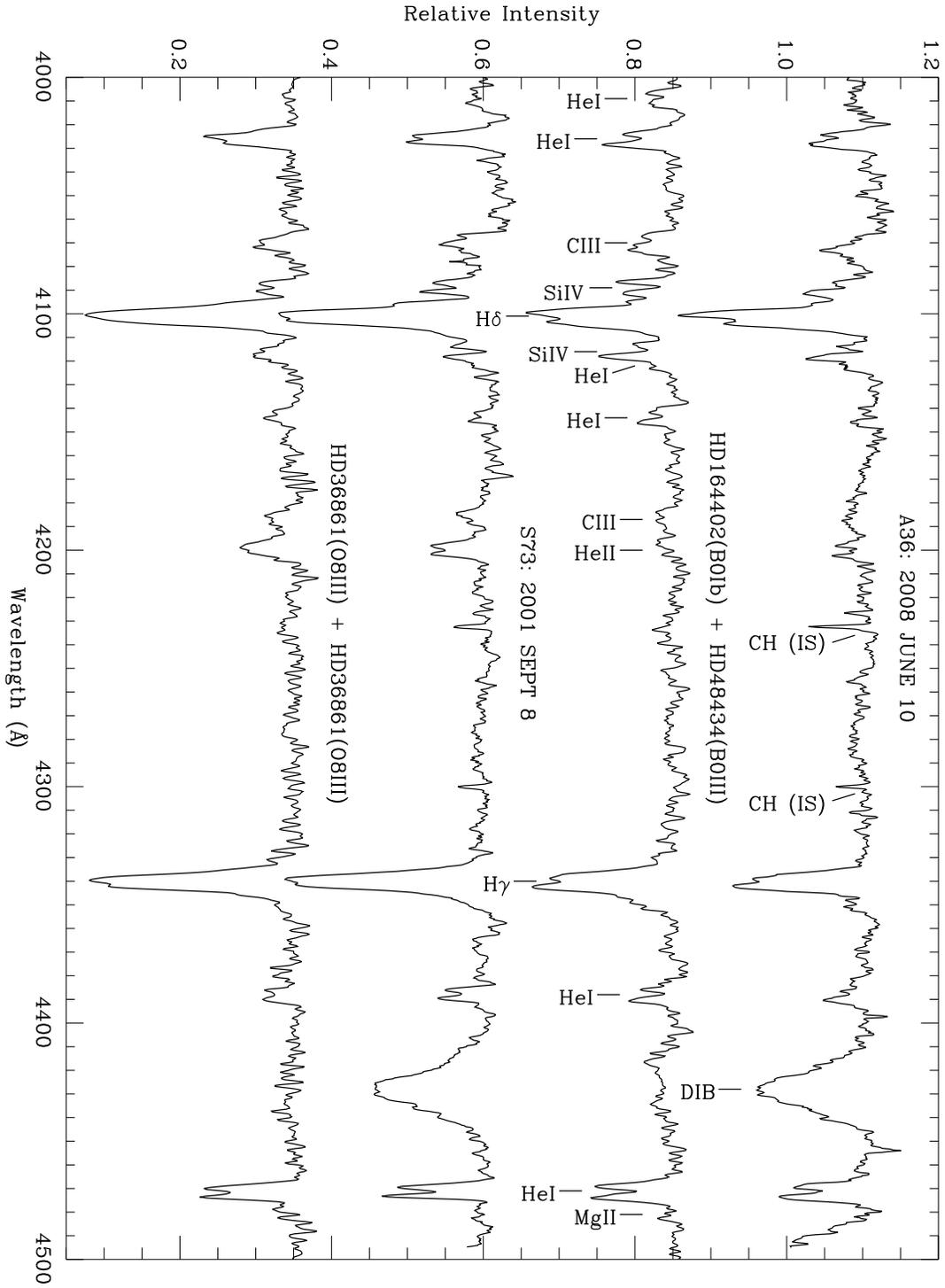}
\caption{Spectra of A36 (\textit{first}) and Schulte~73
(\textit{third})obtained with WIYN on 2008 June 10 and 2001 September
8 respectively and composites of two spectra from the \citet{WF90}
digital atlas shifted to the appropriate radial velocities
(\textit{second and fourth}) that best match the spectral types of the
primaries and secondaries of A36 and Schulte~73. Absorption features
are labeled at their respective intrinsic values.
\label{threeplot}}
\end{figure}

An apparent weakening of the primary's \ion{He}{1} lines between
phases 0.218 and 0.609, seen in panels 1 and 3 of the time series
presented in Figure~\ref{A36progress}, at least partially explains the
large residuals at these epochs and may originate from a Struve-Sahade
effect \citep{Linder07}. This apparent weakening also coincides with
an absence of well-defined absorption profiles in H$\alpha$ (panel 2
of Figure~\ref{A36progress}) between secondary and primary eclipse.
These spectral phenomena coupled with the continuous variation of the
light curve presented in \citet{NSVSa} associated with Algol eclipsing
systems, an O'Connell effect \citep[which is generally associated with
contact or near contact systems;][]{Liu03} and a seemly close
separation between components ($a$sin$i\sim28$~\rsun), implies that
A36 is either a contact or near-contact system. In the very least, it
is likely that one or both components are filling their Roche
lobes. Because A36 is composed of B stars, one may question whether
this is a Be binary. The system does not hold the standard
qualifications for a Be binary however, as the mass ratio is much
greater than 0.1 and the primary is evolved
\citep{Harmanec02}. \citet[][Figure~3]{Conti74} also show that hot
supergiants can have hydrogen emission that very closely resembles the
double-peaked hydrogen emission seen in Be stars.

Is A36 a member of Cyg~OB2? \citet{Hanson03} places doubt on its
status as a member of the cluster based on its disagreement with the
2~Myr isochrone and evolutionary tracks best fit for Cyg OB2
stars. With a systemic velocity offset $-36$~\kms\ from the mean
systemic velocity for cluster stars ($-10.3$~\kms; Paper I), A36 is
likely either a runaway binary system or a background system
masquerading as a member of Cyg OB2. \citet{mcswain07} found
observational evidence for several runaway binary systems, but
concluded that they are rare. These systems can be produced by an
asymmetric supernova of one of the components (producing an SB1) or by
dynamic interactions within a dense cluster. With a stellar density of
$\rho_{0}=40-150$~\msun\ \citep{Kn00} it is certainly possible for A36
to have originated this way, but not probable. However, as a
background object, A36's systemic velocity inferred from a Galactic
rotation curve, would place it at a distance of 6.6 kpc, an interarm
region where OB stars are unlikely to form. Therefore, without a
proper motion to trace its trajectory, or until a complete solution is
computed for the photometric and spectroscopic solutions for this
star, the status of A36's membership will remain uncertain.

\subsection{The SB2 A45}
We obtained 24 observations from 2007 October through 2008 September
on this double-lined system. \citet{Hanson03} classified this star a
B0.5V, but, as with A36, she also suggests it may be an SB2 based on a
high SNR spectrum \citep[see Figure~1 in][]{Hanson03}. We present the
first confirmation of the binary nature of A45 with a spectroscopic
solution to this system.

Of the 24 observations, we used 21 to obtain rough $V_r$ measurements
with the H$\alpha$~$\lambda$6562.80 line. We did not use the
\ion{He}{1}~$\lambda\lambda$5875.75, 6678.15 lines owing to either the
weaker \ion{He}{1} absorption of the secondary or low SNR of the
spectra near \ion{He}{1} $\lambda$5876. The 18 measurements produced a
CLEANed power spectrum with a period estimate of 2.885~days and
aliases of 2~days and 1.55~days.  The latter is a $1-\nu$ alias (where
$\nu$ is the signal frequency per day). Neither alias produced a
folded $V_r$ curve with as low of an rms as 2.885 days. The best
solution for the primary results in a period of $P=2.884\pm0.001$ and
eccentricity of $e=0.273\pm0.002$. The period, eccentricity, and epoch
of periastron were applied as fixed parameters to obtain the solution
for the secondary. The rms of 47~\kms\ indicates there is
room for improvement in the secondary's solution and also explains the
high uncertainty on the calculated masses, $M_1$sin$^3i=12\pm1$~\msun\ and
$M_2$sin$^3i=5.3\pm0.5$~\msun\ ($q=0.6\pm0.1$). With minimum
semi-amplitudes of $K_1=126.1\pm0.3$ \kms\ and $K_2=273\pm17$ \kms\
for the primary and secondary respectively, the solution also produces
minimum semi-major axes of $a_1$sin~$i=6.91\pm0.02$~\rsun\ and
$a_2$sin~$i=14.96\pm0.01$~\rsun\ and a systemic velocity of
$\gamma=5\pm1$~\kms.

We have limited information for classification of the components of
A45 owing to the limited spectral coverage of the WIRO-Longslit
datasets ($\sim$5300--6700~\AA). No attempt to obtain exposures of
this system was made during the 2008 WIYN run because of the system's
angular separation from the rest of the sample. Examination of the
existing spectral coverage and the spectrum in Figure~1 of
\citet{Hanson03} suggests that her classification is more or less
accurate in the sense that both components appear to be early B
stars. There is no evidence of \ion{He}{2} in either component,
\ion{He}{1} is moderately strong, \ion{Mg}{2} is weak, and there
appears to be no \ion{Si}{4}. An absence of the numerous additional
metal lines, such as \ion{O}{2}, \ion{C}{2}, \ion{C}{3}, and
\ion{N}{2} seen in evolved B stars suggests that neither component is
very evolved and both are likely still dwarves. Additionally, a
double-lined system with a mass ratio of $q=0.6\pm0.1$ restricts the
secondary component to within one or two temperature classes of the
primary component. Therefore, we adopt the classification of B0.5V for
the primary and B2V?--B3V? for the secondary component based on the
aforementioned line strengths and mass ratio.

Is A45 a member of Cyg~OB2? Although A45 has the largest angular
separation from the core of Cyg OB2, the systemic velocity agrees with
the mean systemic velocity of cluster members and the system does show
the same diffuse interstellar band features characteristic of Cyg~OB2
stars. Additionally, though the combined luminosity and temperature of
the system disagrees slightly with its predicted placement on the H-R
diagram presented for Cyg~OB2 stars in Figure~12 in \citet{Hanson03},
individual component luminosities and temperatures shift A45's
position down on the diagram and into better agreement with other
Cyg~OB2 members.

\subsection{The SB2 Schulte 73}
Schulte~73 is included in our original sample but excluded from
Paper I owing to the less than 3 observations then
obtained. Spectra taken during the 2008 June WIYN run, however,
revealed the double-lined nature of this star, and re-examination of
the 2001 September 8 spectrum confirmed it. The successive 2008 WIRO
runs cemented its status as a double-lined spectroscopic binary with a
period of $\sim$17.3~days and a mass ratio nearly equal to unity.  The
solution computed for Schulte~73 utilized 27 of the 36 spectra
obtained. Seven spectra from WIYN in 2008 June were excluded for low
SNR owing possibly to a fiber position discrepancy.  Two WIRO spectra
(2008 July 26 \&\ 27) were excluded because of the extremely blended
state of the components.

For the 27 higher SNR spectra, the initial period estimate from the
CLEANed power spectrum is $\sim$16.6~days with an alias at
$\sim$66~days. We dismiss the alias on the basis that progression of
the absorption lines from a blended to deblended state is observed
over approximately 4--5 nights. The best orbital solutions yield a
period of $P=17.28\pm0.03$~days for both components. The exact period,
eccentricity, and date of periastron are held constant for the primary
in order to facilitate a matching solution with the
secondary. Coincidentally, this also provides a better-fitting
solution to the primary $V_r$ curve. We attribute the larger error on
the period (a magnitude higher than the other three systems) to the
smaller number of velocity measurements obtained around apastron
($\phi\simeq0.3-0.6$). The solutions and corresponding $V_r$ curves
for Schulte~73 are shown in Figure~\ref{S73curve}. The filled and open
symbols correspond to the primary and secondary $V_r$ measurements
respectively. The errant WIYN measurements in the primary solution
(triangles) reflect the lower SNR of the 2008 June dataset and the
error in the computed period. With an eccentricity of
$e=0.169\pm0.009$ and semi-amplitudes of $K_1=117\pm2$~\kms\ and
$K_2=117.9\pm0.8$~\kms\, the solutions produce masses of
$M_{1}$sin$^3i=11.2\pm0.2$~\msun\ and
$M_{2}$sin$^3i=11.1\pm0.2$~\msun. This results in a mass ratio of
$q=0.99\pm0.02$ and our first confirmed ``twin'' in the cluster. The
resulting semi-major axes are $a_{1}$sin~$i=39.4\pm0.5$~\rsun\ and
$a_{2}$sin~$i=39.675\pm0.004$~\rsun\ (i.e., a separation of
$a$sin~$i=79.075\pm0.5$~\rsun), and the calculated systemic velocity
is $\gamma=-11\pm2$~\kms.

\citet{MT91} classify this star as an O8V. The components each show
spectral signatures indicative of a later O type star, namely moderate
to weak \ion{He}{2} and stronger \ion{He}{1} absorption. The primary
classification criteria for later O type stars is \ion{He}{2}
$\lambda$4542:\ion{He}{1} $\lambda$4387 and \ion{He}{2}
$\lambda$4200:\ion{He}{1} $\lambda$4144 \citep{WF90}.  The
$\lambda$4200:\ion{He}{1} $\lambda$4144 ratio (also sensitive
to luminosity class) and a visual comparison with the 2001 September 8
spectrum from WIYN seen second from bottom in Figure~\ref{threeplot}
agree best with an O8III spectral type for both components. The
Schulte~73 spectrum has been smoothed to approximately the resolution
of the \citet{WF90} digital atlas. However, without a separation of the
spectral components (e.g., tomographic separation), there is
uncertainty in this assessment. The components also closely match an
O7III and O9III spectral type. The very weak \ion{He}{1}
$\lambda$4144 however, precludes a later temperature class. The
O8III (HD36861) and O8III composite spectrum is shown at the bottom of
Figure~\ref{threeplot}. We also tried various combinations of an O8V,
O8III and O8II. These did not fit the data as well and left larger
residuals when subtracted from the Schulte~73 spectrum.

\section{Partial Solution to the SB1 MT372}
MT372 is a single-lined spectroscopic binary and an eclipsing system,
likely of Algol type, with a period of $P\sim2.228$~days. With a
visual magnitude of nearly 15, only 5 of our 21 spectroscopic
observations of this B0V star are of sufficient SNR for meaningful
analysis. Therefore, to assist in the period determination, we acquired
the available photometric data for MT372 from the public data release
of the Northern Sky Variability Survey \citep{Wozniak04}, and ran it
through our CLEAN IDL program to search for a period. While excluding
points with errors greater than 0.2 magnitudes, we computed a period
of $P=1.114$~days from the power spectrum. In the case that the light
curve exhibits only a primary eclipse, the peak in the power spectrum
corresponds to the real period of the system. In the case that the
light curve exhibits both a primary and secondary eclipse, the peak
corresponds to half the real period. We adopt the latter case because
a period of 1.114~days produces an radial velocity orbital solution
with an unacceptably large rms (where the radial velocities are
determined from the $H\alpha$ line in a similar manner to the SB2s). A
period of $P=2.228$ produces the folded $V_r$ curve and solution shown
in Figure~\ref{MT372curve}. In order to obtain this solution, we fixed
the period and the eccentricity ($e=0$). Allowing the period and/or
eccentricity to vary did not produce a viable solution, and we did not
expect it would because of the very small number of radial velocities
to work with. Although very noisy, this new period also produced a
light curve with two eclipses (primary and secondary; seen in
Figure~\ref{MT372phot}) as opposed to the single primary eclipse
present at a folding of 1.114~days. It should be noted, that the
period of 1.114~days also produced a single eclipse that was
asymmetric, suggesting that the time between primary and secondary
eclipses and secondary and primary eclipses is not equal (i.e., the
system seems to have a slight eccentricity). The orbital solution
shown in Figure~\ref{MT372curve}, also produces a semi-amplitude of
$K_{1}=75\pm20$~\kms, and a semi-major axis of $a_{1}=3.4\pm0.9$.

As this system has little evidence for a double-lined nature, we stay
with the previous spectral classification of this star, B0V. A small
change in velocity width seems to be visible in $H\alpha$ in
Figure~\ref{MT372series} (panel 2), but the SNR of the spectra
(especially evident with the \ion{He}{1} 5876 and 6678 lines shown in
panels 1 \&\ 3 of Figure~\ref{MT372series}) makes it difficult to
produce an accurate determination. The minimum secondary mass
determined from the solution is $M_{2}=3.5\pm0.9$, assuming a primary
of $M_{1}=17.5$ \citep{drilling}. Interpolating from \citet{drilling}
and reasonably assuming that the secondary is also unevolved, this
results in a minimum spectral class of B8V. However, a B8V is unlikely
to be the companion spectral class as it has too low of an effective
temperature to produce the light curve in Figure~\ref{MT372phot}. With
a primary effective temperature of $T_{eff}=30,000$~K \citep{drilling}
and a period of $P=2.228$~days, a B2V at a relatively high inclination
($\sim$75 degrees) would have the correct effective temperature and
stellar radius to produce a light curve that resembles the one for
MT372.  However, given a B2V companion and a mass ratio of $q\sim0.6$,
the computed semi-major axis at approximately 75~degrees is not large
enough to fully separate the two components. This may be explained by
returning to the possible velocity width variation present in the
H$\alpha$ lines. If indeed there is a small amount of blending of
component profiles present, then fitting a single Gaussian profile to
the blended H$\alpha$ line will systematically produce a smaller
primary velocity amplitude and, in turn, smaller semi-amplitude and
semi-major axis. We thus report the listed primary velocity
semi-amplitude as a lower limit.

\section{Summary of Current Survey Results}
In this work, the third part of our ongoing survey to determine the
distribution of binary orbital parameters in the Cyg OB2 association,
we presented three new binary systems (MT145, Schulte~73, \&\ MT372)
and confirmed the binary status of one massive binary candidate
(A45). We also provided the first spectroscopic orbital solution to
the eclipsing binary, A36 and a partial solution to the eclipsing,
single-lined system, MT372. While MT145, Schulte~73, A45, and MT372
are likely members of the cluster based on their systemic velocities
(deviating by only $\sim$0.6--15 \kms\ from the cluster mean of
$-10.3$ \kms\ obtained in Paper I and well within the Association's
radial velocity dispersion) the status of A36, with a systemic
velocity of $\gamma=-47$~\kms, remains uncertain. The system could be
a runaway binary or a background object (though both do not seem very
probable). However, if we do include A36, the current total for OB
binaries in Cyg OB2 is now 18. Of these 18, 14 have period and mass
ratio estimates. In addition, with six partial spectroscopic
solutions, eight full spectroscopic solutions, three partial
photometric solutions, and four full photometric solutions, Cyg OB2 is
on a track to become the cluster with the highest number of OB binary
solutions. At seven it currently ties NGC 6231 \citep{Sana08} for the
highest number of full O-star binary solutions. The locations of all
18 systems are shown in Figure~\ref{binaries}. No evidence of grouping
is apparent.

Table~\ref{Binaries} lists several key parameters for all 18 systems,
including the star designation, photometric and/or spectroscopic
binary type (\textit{Type}), spectral classifications (\textit{S.C.}),
period (\textit{P}), mass ratio when available (\textit{q}), and
literature references (\textit{Ref}).  With the exception of Schulte~9
($P=2.355$~years), periods for all systems are less than 25.2~days,
with a mean of 9.6~days. Eleven systems have a period less than
5~days. 

The 17 systems with periods less than 25.2~days also appear to obey
\"{O}pik's Law \citep{Opik24}, which states that the distribution of
$log(P)$ is flat. To avoid histogram binning selection effects, we
choose to show the normalized cumulative distribution of the logarithm
of the periods (open diamonds) in Figure~\ref{Oepik}. The flat
relation of \"{O}pik's law translates to a linear relation in a
cumulative distribution. For comparison purposes, we also include the
normalized cumulative distributions of O binaries in NGC 6231
\citep[][solid triangles]{Sana08} and the 54 OB binaries from the 9th
Spectroscopic Binary catalog \citep[][asterisks]{Pourbaix04}. All 3
cumulative distributions contain systems with periods less than
25.2~days and were normalized at a period of 8.9~days. To test the
linearity of the Cyg OB2 distribution, we computed its linear Pearson
correlation coefficient. The distribution yields a coefficient of
$r=0.97$ (where $r=1$ is a perfectly linear relation).  The linearity
is illustrated by the best fit line (solid line) in
Figure~\ref{Oepik}. The dashed line in the figure represents a perfect
\"{O}pik distribution (i.e., a perfectly flat distribution within the
range of periods in Cyg OB2). A two sided Kolmogorov-Smirnov ($K-S$)
test between the \"{O}pik and Cyg~OB2 distributions gives a 38\%
probability that both were drawn from the same parent
distribution. $K-S$ tests between Cyg OB2 and NGC 6231 on the one hand
and Cyg OB2 and the 9th Spectroscopic Binary catalog ($P<26$~days) on
the other, yield 91\%\ and 62\%\ probabilities of being drawn from the
same parent distribution, respectively.  Interestingly, the cumulative
distributions of Cyg OB2 and NGC 6231 have a high correlation and are
very similar in appearance, including an upturn near
$log(P)\sim0.4-0.9$. One wonders if this upturn is a true deviation
from a linear relation (i.e., a preference for periods around 5~days),
just a coincidence related to small number statistics, or
observational bias. Though the latter case is a possible explanation
for the upturn in NGC~6231, we do not believe this is the case with
Cyg OB2 because several of the systems with periods around
5~days come from other programs with different observational
cadences (e.g., MT421 and B17).

Figure~\ref{pvse} displays the relation between the orbital periods
and eccentricities for eight of the systems in Cyg OB2 (open
diamonds), six of the systems in NGC~6231 (solid triangles) and 66 of
the systems in the 9th Spectroscopic Binary catalog (asterisks). The
eccentricities in Cyg OB2 range from circular to moderately eccentric
(0.291), while the total distribution in Figure~\ref{pvse} has an
eccentricity range of zero to $\sim$0.75 for periods ranging from
$\sim$0.5~days to 100~days. Systems with periods less than a few days
tend toward more circular orbits, while systems with larger periods
($>10$~days) display a full range of eccentricities. This agrees with
previous studies such as \citet{Giuricin84}. There is a paucity of
highly eccentric systems longer than 30~days. This likely stems from
an incompleteness of the surveys at longer periods and is not a true
trend. The Cyg OB2 star, A45, falls on the high-eccentricity side of
the shorter period systems with an eccentricity of 0.273 and a period
of 2.884~days. It does not appear to be particularly close to the
majority of OB stars in the cluster center (see
Figure~\ref{binaries}), and so dynamical interaction with a massive
neighbor does not seem a likely source of its higher eccentricity
(unless it is a runaway system). It is possible that further
observations of this system will constrain the eccentricity to a
slightly more circular orbit, but its position in the figure is hardly
removed from the rest, and therefore, its moderately elliptical orbit
is not extraordinary. The remainder of Cyg OB2 stars fall well within
the remaining distribution.

With the exception of MT421 and MT429, all the Cyg OB2 systems
have early-type companions and mass ratios widely ranging from
$q\ge0.19$ (MT258) to $q=0.99\pm0.02$ (Schulte~73). One may see this
as an indication that the companions are not randomly drawn from a
\citet{Salpeter55} or similar field star initial mass function
(IMF). However, given that the current sample encompasses a number of
systems that are double-lined spectroscopic binaries (i.e., they are
easier to detect) this assessment may be premature and biased. Only
time will tell, as we have a number of single-lined spectroscopic
binary candidates still to investigate and a spectral database that
has not been completely tapped. We continue to analyze these
candidates with the long term goal of providing accurate distributions
for binary orbital parameters of massive stars with an assumed shared
formation history.

\acknowledgements We thank the time allocation committees of the Lick,
Keck, WIYN, and WIRO observatories for granting us observing time and
making this project possible.  We are also grateful for support from
the National Science Foundation through Research Experience for
Undergraduates (REU) program grant AST 03-53760 and through grant AST
03-07778, and the support of the Wyoming NASA Space Grant Consortium
through grant NNG05G165H. We would also like to graciously thank, in
no particular order, Christopher Rodgers, Emily May, Megan Bagley,
Michael DiPompeo, Michael Alexander, Jessie Runnoe, and Sabrina Cales
for their generous help in observing at WIRO through good skies and
bad, and Jerry Bucher and James Weger for their ample help and upkeep
of WIRO facilities.

\textit{Facilities:} \facility{WIRO ()}, \facility{WIYN ()},
\facility{Lick ()}, \facility{Keck:I ()}

\clearpage

\begin{deluxetable}{lccccc}
\tabletypesize{\scriptsize}
\tablecaption{Summary of Observing Runs \label{obs.tab}}
\tablewidth{0pt}
\tablehead{
\colhead{} &
\colhead{} &
\colhead{} &
\colhead{} &
\colhead{Mean Spectral} &
\colhead{} \\
\colhead{} &
\colhead{Observatory/} &
\colhead{Spectral Coverage} & 
\colhead{Grating} &
\colhead{Resolution} & 
\colhead{Date Coverage} \\
\colhead{Date} &
\colhead{Instrument} &
\colhead{(\AA)} &
\colhead{(l mm$^{-1}$)} &
\colhead{(\AA)} &
\colhead{(HJD)}}

\startdata
1999 Jul 4--5              & Keck/HIRES      & 3890--6270 in 35 orders & 31.6   & 0.1 & 2,451,363--2,451,364 \\
1999 Jul 21--23            & Lick/Hamilton   & 3650--7675 in 81 orders & 31.6   & 0.1 & 2,451,381--2,451,383 \\
1999 Aug 21--23            & Lick/Hamilton   & 3650--7675 in 81 orders & 31.6   & 0.1 & 2,451,411--2,451,413 \\
1999 Oct 14--15            & Keck/HIRES      & 3700--5250 in 29 orders & 31.6   & 0.1 & 2,451,466--2,451,467 \\
2000 Jul 10--11            & Lick/Hamilton   & 3650--7675 in 81 orders & 31.6   & 0.1 & 2,451,736--2,451,737 \\
2000 Sep 18--19            & Keck/HIRES      & 3700--5250 in 29 orders & 31.6   & 0.1 & 2,451,805--2,451,806 \\
2001 Aug 24                & WIYN/Hydra      & 3800--4490 in order 2   & 1200   & 0.9 & 2,452,146            \\
2001 Sep 8--9              & WIYN/Hydra      & 3800--4490 in order 2   & 1200   & 0.9 & 2,452,161--2,452,162 \\
2004 Nov 28--30            & WIYN/Hydra      & 3800--4490 in order 2   & 1200   & 0.9 & 2,453,338--2,453,340 \\
2005 Jul 18--21            & WIRO/WIRO-Spec  & 3800--4490 in order 1   & 2400   & 2.5 & 2,453,570--2,453,573 \\
2005 Jul 18--20,22         & WIRO/WIRO-Spec  & 3800--4490 in order 1   & 2400   & 2.5 & 2,453,632--2,453,635 \\
2005 Oct 13                & WIRO/Longslit   & 4050--6050 in order 2   & 600    & 2.5 & 2,453,657 \\
2006 Jun 16--20            & WIRO/Longslit   & 3900--5900 in order 2   & 600    & 2.5 & 2,453,903--2,453,907 \\
2006 Jul 15--16,20         & WIRO/Longslit   & 3900--5900 in order 2   & 600    & 2.5 & 2,453,932--2,453,935 \\
2006 Sep 8--11             & WIYN/Hydra      & 3800--4490 in order 2   & 1200   & 0.9 & 2,453,987--2,453,990 \\
2006 Oct 7                 & WIRO/Longslit   & 3900--5900 in order 2   & 600    & 2.5 & 2,454,016 \\
2007 Jun 28--30            & WIRO/Longslit   & 3900--5900 in order 2   & 600    & 2.5 & 2,454,280--2,454,282 \\
2007 Jul 4--6              & WIYN/Hydra      & 3820--4510 in order 2   & 1200   & 0.9 & 2,454,285--2,454,287 \\
2007 Aug 28--Sep 4         & WIRO/Longslit   & 5550--6850 in order 1   & 1800   & 1.5 & 2,454,341--2,454,348 \\
2007 Oct 23,25,27--29,31   & WIRO/Longslit   & 5550--6850 in order 1   & 1800   & 1.5 & 2,454,397--2,454,405 \\
2007 Nov 1,3--5            & WIRO/Longslit   & 5550--6850 in order 1   & 1800   & 1.5 & 2,454,406--2,454,410 \\
2008 Jun 10--15            & WIYN/Hydra      & 3820--4510 in order 2   & 1200   & 0.9 & 2,454,628--2,454,633 \\
2008 Jun 23--30            & WIRO/Longslit   & 5210--6680 in order 1   & 1800   & 1.5 & 2,454,641--2,454,648 \\
2008 Jul 21,22--27         & WIRO/Longslit   & 5250--6740 in order 1   & 1800   & 1.5 & 2,454,669--2,454,675 \\
2008 Aug 17--22            & WIRO/Longslit   & 5250--6740 in order 1   & 1800   & 1.5 & 2,454,696--2,454,701 \\
2008 Sep 12,13--16,19      & WIRO/Longslit   & 5250--6740 in order 1   & 1800   & 1.5 & 2,454,722--2,454,729 \\
2008 Oct 14--18            & WIRO/Longslit   & 5250--6740 in order 1   & 1800   & 1.5 & 2,454,754--2,454,758 \\
\enddata
\end{deluxetable}

\clearpage

\pagestyle{plaintop}

\begin{deluxetable}{
lrrrrr}
\tabletypesize{\tiny}
\tabletypesize{\scriptsize}
\tablewidth{0pt}
\tablecaption{Orbital Elements \label{orbparms.tab}}
\tablehead{
\colhead{Element} &
\colhead{MT145} & 
\colhead{A36} & 
\colhead{A45} &
\colhead{Schulte~73} &
\colhead{MT372}}  
\startdata
$P$ (Days)             & 25.140 (0.008)              & 4.674 (0.004)     & 2.884 (0.001)      & 17.28 (0.03)    & 2.228 (fixed) \\
$e$                    & 0.291  (0.009)              & 0.10 (0.01)       & 0.273 (0.002)      & 0.169 (0.009)   & 0.0 (fixed)   \\
$\omega$ (deg)         & 158.3 (0.6)                 & 359.2 (0.7)       & 188.8 (0.1)        & 7.3 (0.8)       & 344 (146)     \\
$\gamma$ (\kms)        & -19.3 (0.4)                 & -47 (2)           & 5 (1)              & -11 (2)         & 3 (7)         \\
$T_0$ (HJD-2,400,000)  & 51789.08 (0.04)             & 54693.836 (0.009) & 54728.209 (0.001)  & 54700.89 (0.03) & 54755 (9)     \\
$K_{1}$ (\kms)         & 48.9 (0.7)                  & 169 (9)           & 126.1 (0.3)        & 117 (2)         & 75 (20)       \\
$K_{2}$ (\kms)         & \nodata                     & 243 (2)           & 273 (17)           & 117.9 (0.8)     & \nodata       \\
$M_{1}$ (\msun)        & 22.6 (0.5)\tablenotemark{a} & $>$19.8 (0.9)\tablenotemark{c}     & $>$12 (1)\tablenotemark{c}           & $>$11.2 (0.2)\tablenotemark{c}    & 17.5\tablenotemark{b} \\
$M_{2}$ (\msun)        & 6.0 (0.1)--13.8 (0.6)       & $>$13.8 (0.9)\tablenotemark{c}     & $>$5.3 (0.5)\tablenotemark{c}        & $>$11.1 (0.2)\tablenotemark{c}    & $\sim$10.0\tablenotemark{b} \\
$f(m)_1$ (\msun)       & 0.267 (0.012)               & 6.9 (0.2)         & 0.534 (0.004)      & 2.8 (0.1)       & 0.099 (0.080) \\
$f(m)_2$ (\msun)       & \nodata                     & 2.3 (0.4)         & 5 (1)              & 2.81 (0.06)     & \nodata       \\
S.~C.$_1$              & O9III                       & B0Ib              & B0.5V              & O8III           & B0V           \\
S.~C.$_2$              & mid B                       & B0III             & B2V?--B3V?         & O8III           & B2?V          \\
$a_{1}$sin~$i$ (\rsun) & 23.2 (0.3)                  & 15.6 (0.8)        & 6.91 (0.02)        & 39.4 (0.5)      & 3.4 (0.9)     \\
$a_{2}$sin~$i$ (\rsun) & \nodata                     & 22.345 (0.003)    & 14.96 (0.01)       & 39.675 (0.004)  & \nodata       \\
$rms_1$ (\kms)         & 8.9                         & 29.3              & 25                 & 5.1             & 14.9          \\
$rms_2$ (\kms)         & \nodata                     & 10.0              & 47                 & 6.7             & \nodata       \\

\enddata
\tablecomments{Calculated errors are located in parentheses.}
\tablenotetext{a}{Theoretical O star masses adopted from \citet{FM05}.}
\tablenotetext{b}{Theoretical B star masses interpolated from \citet{drilling}.}
\tablenotetext{c}{Calculated mass is equal to $M$sin$^3i$.}

\end{deluxetable}

\clearpage

\pagestyle{plaintop}

\begin{deluxetable}{ccrrr}
\tabletypesize{\tiny}
\tabletypesize{\scriptsize}
\tablewidth{0pc}
\tablecaption{Ephemerides for MT145 \&\ MT372 \label{OC}}
\tablehead{
\colhead{} & 
\colhead{} & 
\colhead{$V_r$} &
\colhead{1$\sigma$ err} &
\colhead{$O-C$} \\
\colhead{Date (HJD-2,400,000)} & 
\colhead{$\phi$} &
\colhead{(\kms)} &
\colhead{(\kms)} &
\colhead{(\kms)}}
\tablecolumns{5}
\startdata  
51,381.50.......................... & 0.788  &   1.7 & 10.3   &   1.8  \\
51,383.50.......................... & 0.867  & -24.7 & 15.1   &  -1.0  \\
51,467.90.......................... & 0.225  & -21.5 &  4.4   &   9.2  \\
51,736.50.......................... & 0.909  & -33.5 & 11.3   &   7.8  \\
51,737.50.......................... & 0.948  & -56.4 & 13.7   &   3.5  \\
51,805.76.......................... & 0.664  &  22.8 &  3.2   &   7.9  \\
52,146.70.......................... & 0.225  & -23.6 &  3.6   &   6.9  \\
52,161.81.......................... & 0.826  &  -7.4 &  4.1   &   2.4  \\
52,162.67.......................... & 0.860  & -17.8 &  3.8   &   3.3  \\
53,338.63.......................... & 0.637  &  14.0 &  4.4   &  -1.9  \\
53,340.59.......................... & 0.715  &  10.7 &  4.2   &  -0.4  \\
53,570.50.......................... & 0.860  & -40.0 &  8.3   & -18.9  \\
53,572.50.......................... & 0.940  & -66.4 & 10.0   & -10.3  \\
53,632.50.......................... & 0.327  & -10.5 & 11.5   &  -2.7  \\
53,633.50.......................... & 0.367  & -10.6 &  6.7   &  -9.3  \\
53,636.50.......................... & 0.486  &  11.5 &  5.2   &  -0.3  \\
53,657.50.......................... & 0.321  &  -0.8 &  6.8   &   8.0  \\
53,903.75.......................... & 0.116  & -71.7 &  8.4   &  -7.1  \\
53,903.76.......................... & 0.117  & -74.7 &  9.1   & -10.3  \\
53,904.72.......................... & 0.155  & -33.1 &  7.8   &  18.8  \\
53,904.73.......................... & 0.155  & -41.0 &  7.6   &  10.7  \\
53,904.90.......................... & 0.162  & -65.6 &  7.7   & -16.1  \\
53,905.77.......................... & 0.197  & -49.4 &  6.8   & -10.6  \\
53,905.90.......................... & 0.202  & -48.1 &  7.3   & -11.0  \\
53,906.74.......................... & 0.235  & -19.3 &  8.5   &   8.6  \\
53,906.89.......................... & 0.241  & -42.2 & 10.6   & -15.8  \\
53,907.78.......................... & 0.277  & -20.5 & 10.2   &  -2.7  \\
53,907.88.......................... & 0.280  & -28.4 &  7.1   & -11.4  \\
53,932.75.......................... & 0.270  & -17.0 &  7.9   &   2.5  \\
53,932.85.......................... & 0.274  &   2.5 &  8.8   &  20.9  \\
53,932.86.......................... & 0.274  &   6.4 & 10.3   &  24.8  \\
53,935.89.......................... & 0.395  & -14.7 & 12.8   & -17.3  \\
53,987.70.......................... & 0.456  &   1.8 &  4.6   &  -7.6  \\
53,988.75.......................... & 0.497  &   9.9 &  4.4   &  -2.8  \\
53,989.66.......................... & 0.534  &  13.7 &  4.7   &  -1.0  \\
53,990.86.......................... & 0.581  &  17.3 &  4.3   &   1.1  \\
54,280.83.......................... & 0.116  & -61.7 &  7.0   &   3.0  \\
54,280.87.......................... & 0.117  & -56.5 &  6.6   &   7.8  \\
54,281.70.......................... & 0.150  & -60.2 &  8.5   &  -6.7  \\
54,281.78.......................... & 0.153  & -50.9 &  7.8   &   1.4  \\
54,281.85.......................... & 0.156  & -51.5 &  7.0   &  -0.0  \\
54,282.70.......................... & 0.190  & -44.8 &  7.9   &  -4.1  \\
54,282.77.......................... & 0.193  & -45.6 &  6.0   &  -5.7  \\
54,282.83.......................... & 0.195  & -45.9 &  6.9   &  -6.8  \\
54,285.93.......................... & 0.318  &  -6.0 &  4.3   &   3.3  \\
54,286.66.......................... & 0.348  &  -7.6 &  5.2   &  -3.4  \\
54,287.73.......................... & 0.390  &   5.3 &  4.3   &   3.2  \\
54,397.71.......................... & 0.765  &   5.0 & 16.7   &   0.7  \\
54,399.69.......................... & 0.843  & -22.0 & 15.3   &  -6.9  \\
54,401.67.......................... & 0.922  & -47.9 & 16.0   &  -0.3  \\
54,402.71.......................... & 0.964  & -54.1 & 14.6   &  12.3  \\
\cutinhead{MT372}
54,754.62.......................... & 0.949  &  75.4 &  7.4   &   9.8  \\
54,755.60.......................... & 0.388  & -36.6 &  6.3   &   1.7  \\
54,756.60.......................... & 0.840  &  16.0 &  3.0   &  -8.9  \\
54,757.59.......................... & 0.283  &   6.9 &  3.9   &  -1.9  \\
54,758.72.......................... & 0.788  &   6.9 &  3.4   &   6.3  \\
\enddata 

\end{deluxetable}

\clearpage
\pagestyle{plaintop}

\begin{deluxetable}{lcrrrr}
\tabletypesize{\tiny}
\tabletypesize{\scriptsize}
\tablewidth{0pc}
\tablecaption{Ephemerides for A36, A45, \&\ Schulte~73 \label{OC2}}
\tablehead{
\colhead{} & 
\colhead{} & 
\colhead{$V_{r1}$} &
\colhead{$O_1-C_1$} &
\colhead{$V_{r2}$} &
\colhead{$O_2-C_2$} \\ 
\colhead{Date (HJD-2,400,000)} &
\colhead{$\phi$} &
\colhead{(\kms)} &
\colhead{(\kms)} &
\colhead{(\kms)} &
\colhead{(\kms)}}
\tablecolumns{6}
\startdata  
\cutinheadb{A36}
54,403.61..........................  &  0.910 & -163.4 (5.6)  &    27.7 &  167.5 (4.9) &     4.2 \\ 
54,403.70..........................  &  0.928 & -174.9 (5.0)  &    30.9 &  184.7 (4.3) &     1.0 \\ 
54,405.71..........................  &  0.357 &  103.4 (5.3)  &    47.6 & -194.2 (3.5) &     3.1 \\ 
54,406.63..........................  &  0.555 &  105.4 (3.7)  &     5.6 & -261.1 (3.4) &    -5.3 \\ 
54,406.76..........................  &  0.582 &  108.6 (3.7)  &    16.2 & -244.5 (3.7) &     0.2 \\ 
54,409.73..........................  &  0.218 &  -23.6 (5.5)  &    45.3 &  -23.0 (3.7) &    -3.3 \\ 
54,410.67..........................  &  0.419 &   98.4 (3.7)  &    10.5 & -243.8 (3.4) &    -1.8 \\ 
54,628.88..........................  &  0.103 & -163.3 (2.6)  &    24.6 &  147.2 (2.5) &    -5.3 \\ 
54,633.74..........................  &  0.143 & -150.6 (3.1)  &    -0.8 &  108.1 (3.3) &    11.0 \\ 
54,669.84..........................  &  0.867 & -116.5 (6.2)  &    34.0 &  113.4 (4.1) &     7.4 \\ 
54,672.78..........................  &  0.495 &  100.3 (3.8)  &    -4.4 & -280.6 (3.7) &   -16.1 \\ 
54,673.88..........................  &  0.730 &  -37.7 (12.9) &   -33.6 &  -92.2 (7.1) &    11.6 \\ 
54,696.69..........................  &  0.610 &  109.0 (14.4) &    27.6 & -236.8 (3.5) &    -8.4 \\ 
54,696.96..........................  &  0.669 &  125.2 (4.3)  &    79.0 & -155.7 (3.9) &    21.1 \\ 
54,697.66..........................  &  0.818 & -129.9 (5.9)  &   -32.7 &   23.1 (3.4) &    -6.8 \\ 
54,697.96..........................  &  0.882 & -163.1 (4.3)  &     2.4 &  129.6 (3.7) &     2.4 \\ 
54,698.75..........................  &  0.052 & -198.2 (4.4)  &    24.4 &  200.8 (7.1) &    -2.9 \\ 
54,700.87..........................  &  0.505 &  103.7 (3.7)  &    -1.3 & -254.3 (3.4) &    10.3 \\ 
54,701.90..........................  &  0.724 &    1.7 (3.6)  &     0.8 & -114.0 (3.8) &    -2.9 \\ 
\cutinhead{A45}
54,403.66........................... &  0.447 &    106.3 (14.89) &     -6.7 &  -257.5 (5.33)  &  -28.2 \\
54,403.74........................... &  0.476 &    127.4 (12.48) &     14.6 &  -225.1 (6.17)  &    3.9 \\
54,405.72........................... &  0.161 &      7.9 (5.40)  &      3.8 &   -33.0 (25.14) &  -39.6 \\
54,405.74........................... &  0.169 &    -56.9 (6.78)  &    -67.9 &   -12.9 (16.19) &   -4.5 \\
54,406.64........................... &  0.483 &     87.3 (15.39) &    -25.3 &  -200.5 (8.72)  &   28.0 \\
54,406.65........................... &  0.486 &     87.3 (20.36) &    -25.1 &  -165.5 (11.95) &   62.6 \\
54,406.77........................... &  0.527 &    135.4 (13.52) &     26.4 &  -171.1 (7.89)  &   49.6 \\
54,409.74........................... &  0.557 &    126.6 (9.00)  &     21.8 &  -226.0 (5.93)  &  -14.5 \\
54,410.68........................... &  0.882 &    -84.1 (14.51) &     -9.6 &   233.8 (6.10)  &   57.1 \\
54,641.84........................... &  0.046 &   -126.0 (5.87)  &    -14.6 &   226.6 (6.80)  &  -30.0 \\
54,647.81........................... &  0.116 &    -28.5 (8.80)  &     11.4 &     4.0 (8.07)  &  -97.9 \\
54,672.90........................... &  0.817 &      9.7 (15.16) &     26.3 &     8.3 (19.51) &  -43.0 \\
54,724.70........................... &  0.784 &    -17.0 (17.09) &    -26.4 &     4.0 (23.56) &    9.0 \\
54,724.78........................... &  0.810 &     23.8 (7.31)  &     34.1 &    -1.4 (23.90) &  -39.0 \\
54,725.72........................... &  0.135 &    -17.0 (10.74) &      3.9 &     6.8 (8.43)  &  -53.8 \\
54,726.72........................... &  0.483 &    115.3 (10.66) &      2.7 &  -250.6 (7.71)  &  -22.2 \\
54,726.79........................... &  0.508 &    111.6 (13.80) &      0.6 &  -228.2 (6.83)  &   -3.2 \\
54,729.70........................... &  0.517 &    139.9 (11.19) &     29.8 &  -234.4 (6.08)  &  -11.2 \\
54,747.59........................... &  0.721 &     21.5 (5.86)  &    -27.6 &    -4.1 (6.10)  &   87.0 \\
54,748.60........................... &  0.071 &    -76.2 (6.34)  &     11.4 &   285.4 (26.42) &   80.1 \\
54,754.77........................... &  0.212 &    -15.6 (10.59) &    -59.0 &     3.2 (44.57) &   81.7 \\
\cutinhead{Schulte~73}
52,161.65..........................  & 0.071  & -133.5 (2.0) &  -1.9  & 110.7 (2.0) &   1.5   \\
54,628.75..........................  & 0.825  & -118.3 (2.9) & -43.6  &  86.5 (2.5) &  31.3   \\
54,628.88..........................  & 0.833  & -112.4 (3.6) & -31.8  &  89.1 (2.1) &  27.9   \\
54,629.87..........................  & 0.890  & -157.5 (3.1) & -34.0  &  88.6 (2.1) & -15.3   \\
54,630.74..........................  & 0.941  & -156.9 (2.9) &  -4.9  & 121.3 (2.1) & -10.5   \\
54,630.85..........................  & 0.947  & -166.7 (2.5) & -12.3  & 122.3 (2.1) & -11.9   \\
54,633.86..........................  & 0.121  & -115.2 (2.4) & -22.0  &  95.6 (2.1) &  25.5   \\
54,696.75..........................  & 0.760  &  -31.4 (6.8) &  -5.4  &   0.5 (2.0) &  -5.7   \\
54,696.96..........................  & 0.772  &  -13.4 (2.9) &  21.2  &  16.1 (3.8) &   1.1   \\
54,697.77..........................  & 0.819  &  -67.9 (5.3) &   1.9  &  46.8 (1.8) &  -3.5   \\
54,697.85..........................  & 0.824  &  -81.4 (3.0) &  -8.0  &  60.9 (2.0) &   6.9   \\
54,697.92..........................  & 0.828  &  -70.8 (2.9) &   5.9  &  69.9 (2.4) &  12.7   \\
54,698.71..........................  & 0.874  & -122.4 (1.9) & -10.5  &  88.7 (1.8) &  -3.7   \\
54,698.91..........................  & 0.885  & -125.2 (3.1) &  -5.0  & 107.1 (2.8) &   6.5   \\
54,699.72..........................  & 0.932  & -148.7 (1.4) &  -0.6  & 127.3 (1.4) &  -0.8   \\
54,699.93..........................  & 0.944  & -142.4 (1.7) &  10.9  & 138.2 (1.6) &   5.1   \\
54,700.72..........................  & 0.990  & -150.0 (2.1) &  11.8  & 131.3 (1.6) &  -9.5   \\
54,701.73..........................  & 0.048  & -155.0 (1.8) &  -9.8  & 126.9 (1.5) &   3.7   \\
54,701.94..........................  & 0.061  & -142.0 (2.7) &  -3.8  & 124.4 (2.2) &   8.4   \\
54,724.69..........................  & 0.377  &   66.8 (3.3) &   6.6  & -91.7 (2.8) &  -8.2   \\
54,724.81..........................  & 0.384  &   73.1 (1.8) &  11.2  & -88.7 (1.4) &  -3.6   \\
54,725.69..........................  & 0.435  &   69.4 (1.5) &  -1.0  & -93.3 (1.4) &  -0.1   \\
54,725.80..........................  & 0.441  &   70.8 (1.6) &  -0.2  & -93.3 (1.5) &   0.4   \\
54,726.70..........................  & 0.493  &   69.0 (1.7) &  -3.0  & -90.4 (1.6) &   3.9   \\
54,726.82..........................  & 0.500  &   67.5 (1.6) &  -4.2  & -87.9 (1.5) &   5.9   \\
54,729.66..........................  & 0.664  &   27.0 (2.0) &  -2.4  & -52.7 (1.9) &  -2.8   \\
54,729.84..........................  & 0.675  &   22.2 (2.1) &  -2.2  & -52.9 (2.1) &  -8.1   \\
\enddata 

\end{deluxetable}

\clearpage
\thispagestyle{empty}

\begin{deluxetable}{lccccl}
\tabletypesize{\tiny}
\tabletypesize{\scriptsize}
\tablewidth{0pc}
\tablecaption{OB Binaries in Cyg OB2 \label{Binaries}}
\tablehead{
\colhead{} & 
\colhead{} & 
\colhead{} & 
\colhead{P} &
\colhead{} &
\colhead{} \\
\colhead{Star} &
\colhead{Type} &
\colhead{S.C.} &
\colhead{(days)} &
\colhead{q} &
\colhead{Ref.}}
\startdata
MT059      & SB1     & O8V \&\ B                          & 4.8527 (0.0002)      & 0.22--0.67\tablenotemark{a}  & 1 \\
MT145      & SB1     & O9III \&\ mid B                    & 25.140 (0.008)       & 0.26--0.63\tablenotemark{a}  & 2 \\
MT252      & SB2     & B2III \&\ B1V                      & 18--19               & 0.8 (0.2)                    & 1 \\
MT258      & SB1     & O8V \&\ B                          & 14.660 (0.002)       & 0.18--0.89\tablenotemark{a}  & 1 \\
MT372      & EA?/SB1 & B0V \&\ B2?V                       & 2.228 (fixed)        & $\sim$0.6                 	& 2,3 \\
MT421      & EA      & O9V \&\ B9V--A0V                   & 4.161   	         & $\sim$0.16--0.19          	& 4 \\
MT429      & EA      & B0V \&\ ??                         & 2.9788  	         & \nodata                   	& 4 \\
MT696  & SB2/EW/KE   & O9.5V \&\ early B                  & 1.46    	         & \nodata                   	& 5 \\
MT720      & SB2     & early B \&\ early B                & $<$ 5                & \nodata                   	& 1 \\
MT771      & SB2     & O7V \&\ O9V                        & 1.5:                 & 0.8 (0.1)	             	& 1 \\
Schulte 3  & SB2/EA? & O6IV? \&\ O9III                    & 4.7464 (0.0002)      & 0.44 (0.08)               	& 1,6 \\
Schulte 5  & EB      & O7Ianfp \&\ Ofpe/WN9 (\&\ B0V?)    & 6.6                  & 0.28 (0.02)               	& 7,8,9,10,11,12 \\
Schulte 8a & SB2     & O5.5I \&\ O6?                      & 21.908 	         & 0.86 (0.04)               	& 13,14 \\
Schulte 9  & SB2     & O5? \&\ O6--7?                     & 2.355~yr             & \nodata                   	& 15      \\
Schulte 73 & SB2     & O8III \&\ O8III                    & 17.28 (0.03)         & 0.99 (0.02)               	& 2      \\
A36        & SB2/EA  & B0Ib \&\ B0III                     & 4.674 (0.004)        & 0.70 (0.06)               	& 2,16,17   \\
A45        & SB2     & B0.5V \&\ B2V?--B3V?               & 2.884 (0.001)        & 0.46 (0.02)               	& 2,17      \\
B17        & SB2     & O7? \&\ O9?                        & 4.0217 (0.0004)      & $\sim$1?                  	& 18,19     \\
\enddata

\tablecomments{Photometric types EW/KE, EA, and EB stand for Contact
system of the W UMa type (ellipsoidal; $P<1$ day), Algol type (near
spherical), and $\beta$ Lyr type (ellipsoidal; $P>1$ day)
respectively. The mass ratio for MT421 is calculated using the O star
masses of \citet{FM05} and interpolated AB masses of
\citet{drilling}. The mass ratio for Schulte~10 is calculated using
the interpolated B star masses from \citet{drilling}.}
\tablenotetext{a}{The range in mass ratio incorporates both observational and inclination uncertainties}

\tablerefs{
(1) \citet{Kiminki08} 
%MT372,MT145,S73,A36,A45
(2) This work;
(3) \citet{Wozniak04};
%MT421 and MT429
(4) \citet{PJ98}; 
%MT696
(5) \citet{Rios04};
%Schulte 3
(6) \citet[in prep]{Karen}; 
%Schulte 5
(7) \citet{Wilson48}; 
(8) \citet{Wilson51}; 
(9) \citet{Mics53};
(10) \citet{Wal73}; 
(11) \citet{Contreras97}; 
(12) \citet{Rauw99}; 
%Schulte 8a
(13) \citet{Romano69}; 
(14) \citet{Debeck04};
(15) \citet{Naze08};
%A36
(16) \citet{NSVSa};
(17) \citet{Hanson03};
%B17
(18) \citet[][submitted]{Stroud09};
(19) \citet{NSVSb}}

\end{deluxetable}

\clearpage

\begin{figure}
\epsscale{1.0}
\plotone{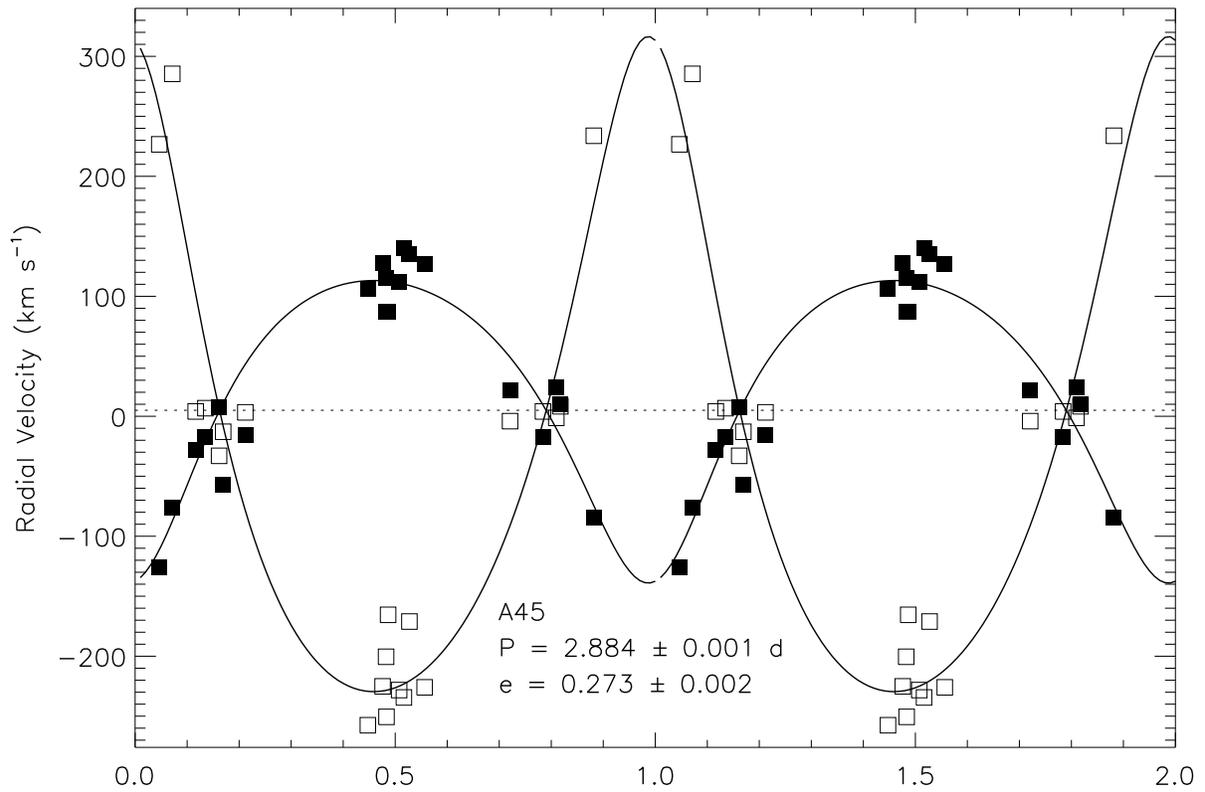}
\caption{$V_r$ curve and orbital solution for A45 using 21 of the
24 observations. The filled points correspond to the primary (B0.5V) and
the open points correspond to the secondary (B2V?--B3V?). All data were obtained
with the WIRO-Longslit spectrograph.
\label{A45curve}}
\end{figure}

\begin{figure}
\epsscale{1.0} \plotone{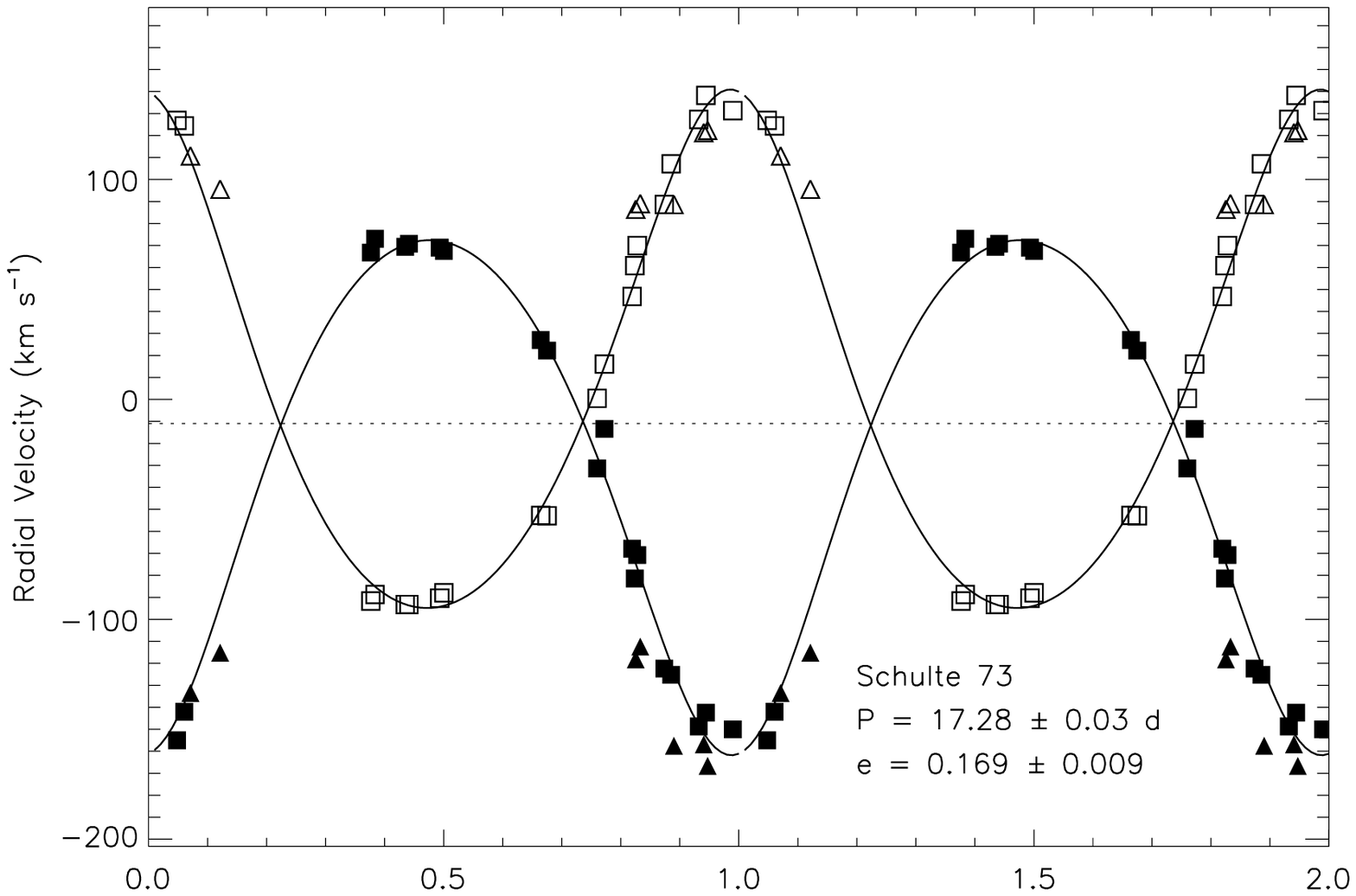}
\caption{$V_r$ curve and orbital solution for S73 using 27 of the
36 observations. The filled points correspond to the primary (O8III) and
the open points correspond to the secondary (O8III). The symbols represent
the same observation locations as in Figure~1.
\label{S73curve}}
\end{figure}

\begin{figure}
\epsscale{1.0} \plotone{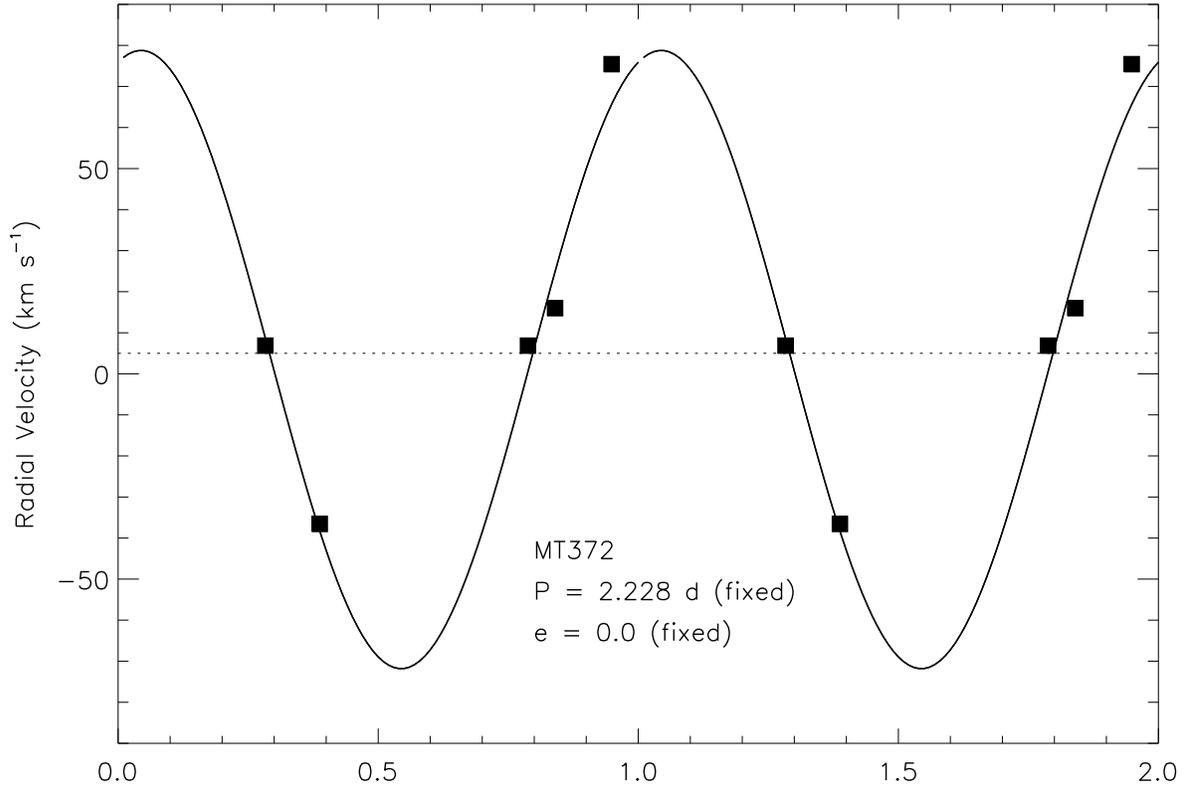}
\caption{$V_r$ curve and orbital solution for MT372 using 5 H$\alpha$
observations and the period obtained from the photometric data. All
data were obtained with the WIRO-Longslit spectrograph.
\label{MT372curve}}
\end{figure}

\begin{figure}
\epsscale{1.0} \plotone{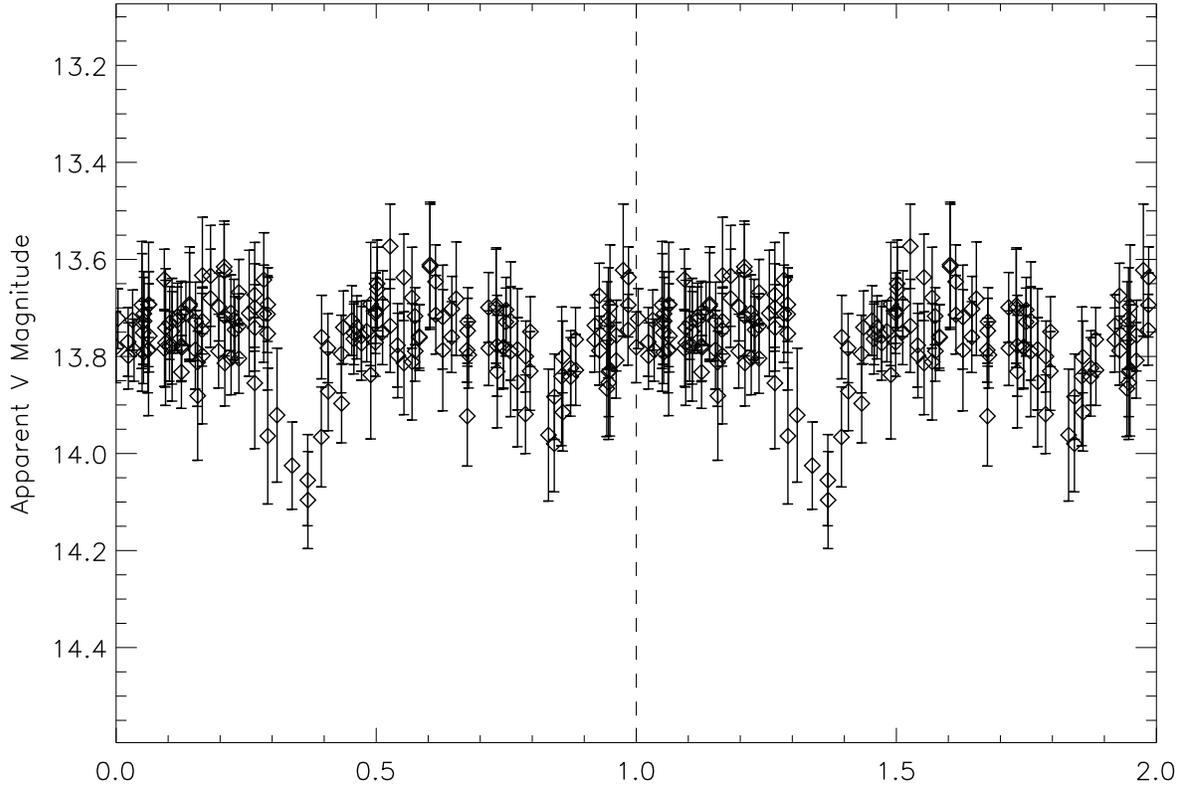}
\caption{Folded light curve of MT372 ($P=2.228$~d) utilizing photometric
data from the Northern Sky Variability Survey
\citep{Wozniak04}.
\label{MT372phot}}
\end{figure}

\begin{figure}
\epsscale{1.0} \plotone{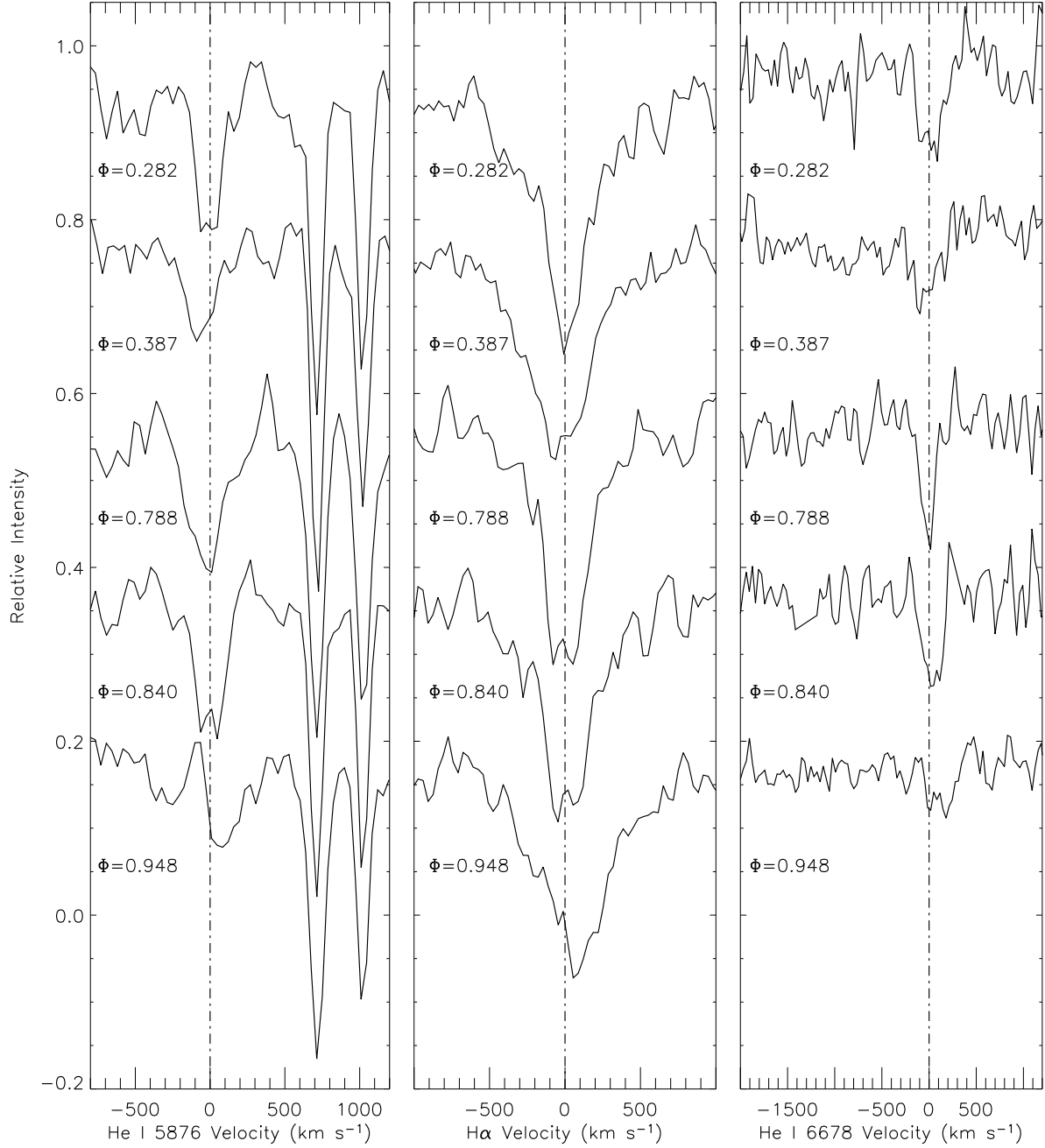}
\caption{\ion{He}{1}~$\lambda$5875.75 (left),
H$\alpha$~$\lambda$6562.80 (middle),
\ion{He}{1}~$\lambda$6678.15 (right) in velocity space and in order of phase for
observations of MT372 taken at WIRO 2008 October 14--18.
\label{MT372series}}
\end{figure}

\begin{figure}
\epsscale{1.0}
\plotone{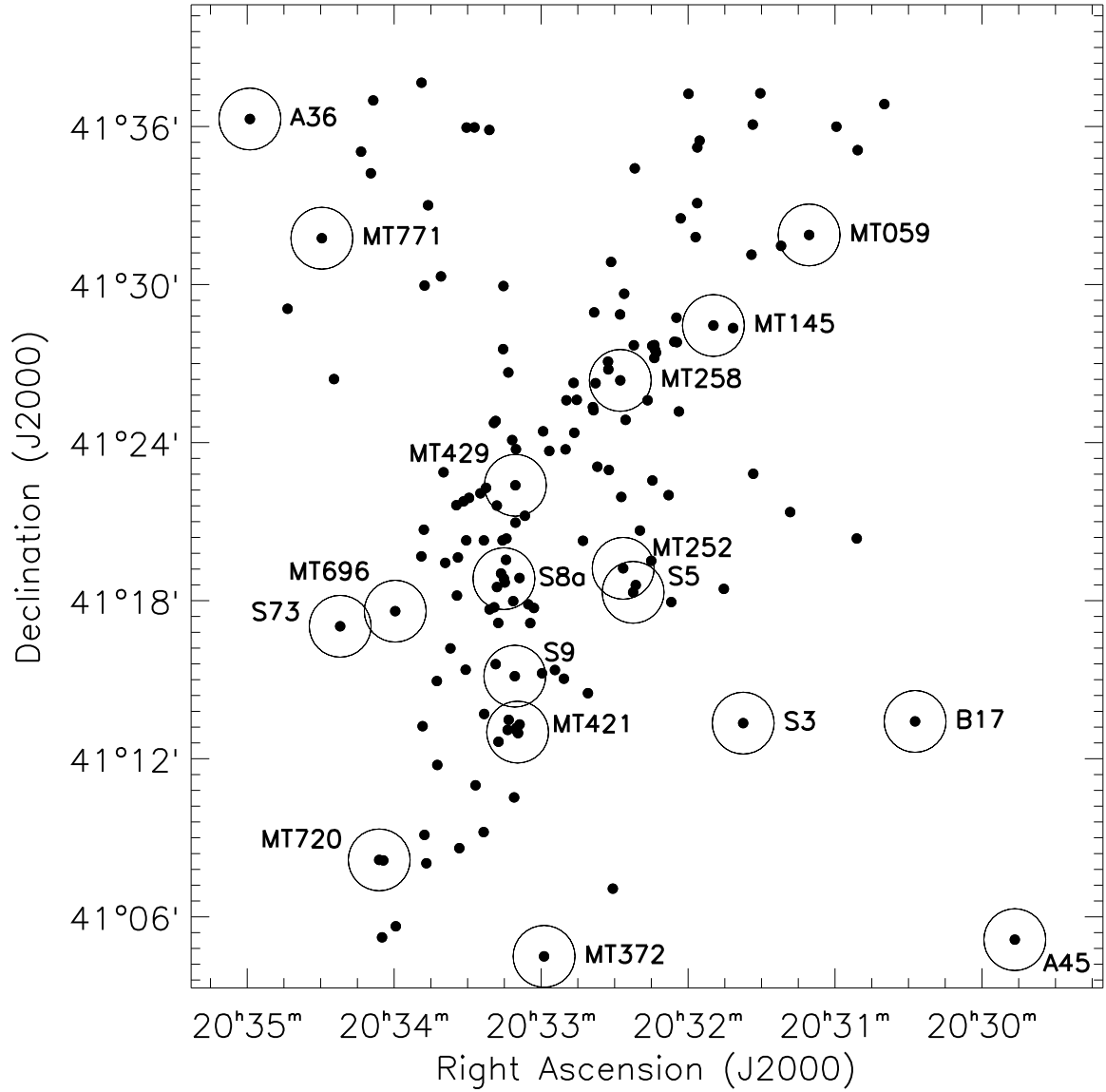}
\caption{A map of OB stars in Cyg OB2. Circles indicate the 
locations of the 18 known binaries within Cyg~OB2. 
\label{binaries}}
\end{figure}

\begin{figure}
\epsscale{1.0}
\plotone{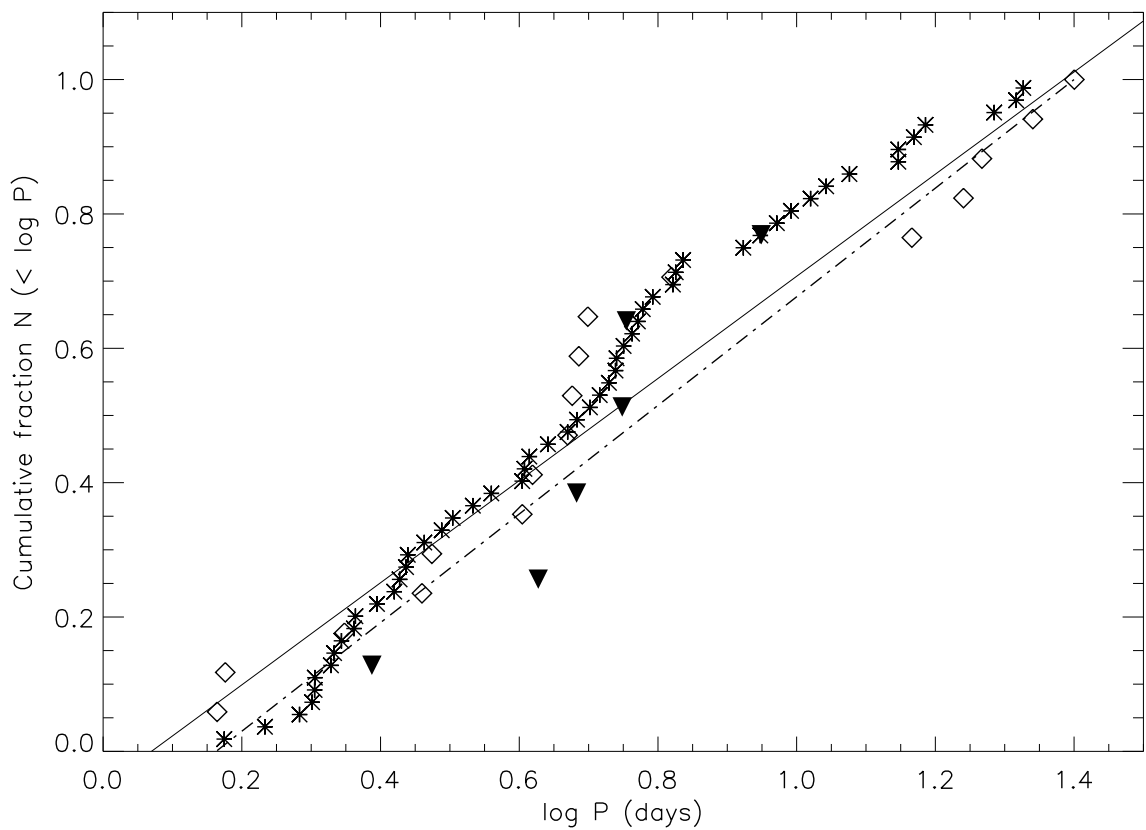}
\caption{Cumulative fractions of the number of binary systems with
periods less than 25.2~days in Cyg OB2 (open diamonds), NGC 6231
(solid triangles), and the OB stars from the 9th Spectroscopic Binary
catalog \citep[][asterisks]{Pourbaix04}, as a function of the
logarithm of the period.  The solid line represents the best linear
fit to the Cyg OB2 distribution and is included for illustrative
purposes. The dashed line represents a perfect \"{O}pik distribution
\citep[i.e., the distribution of log(P) is flat;][]{Opik24}.
\label{Oepik}}
\end{figure}

\begin{figure}
\epsscale{1.0}
\plotone{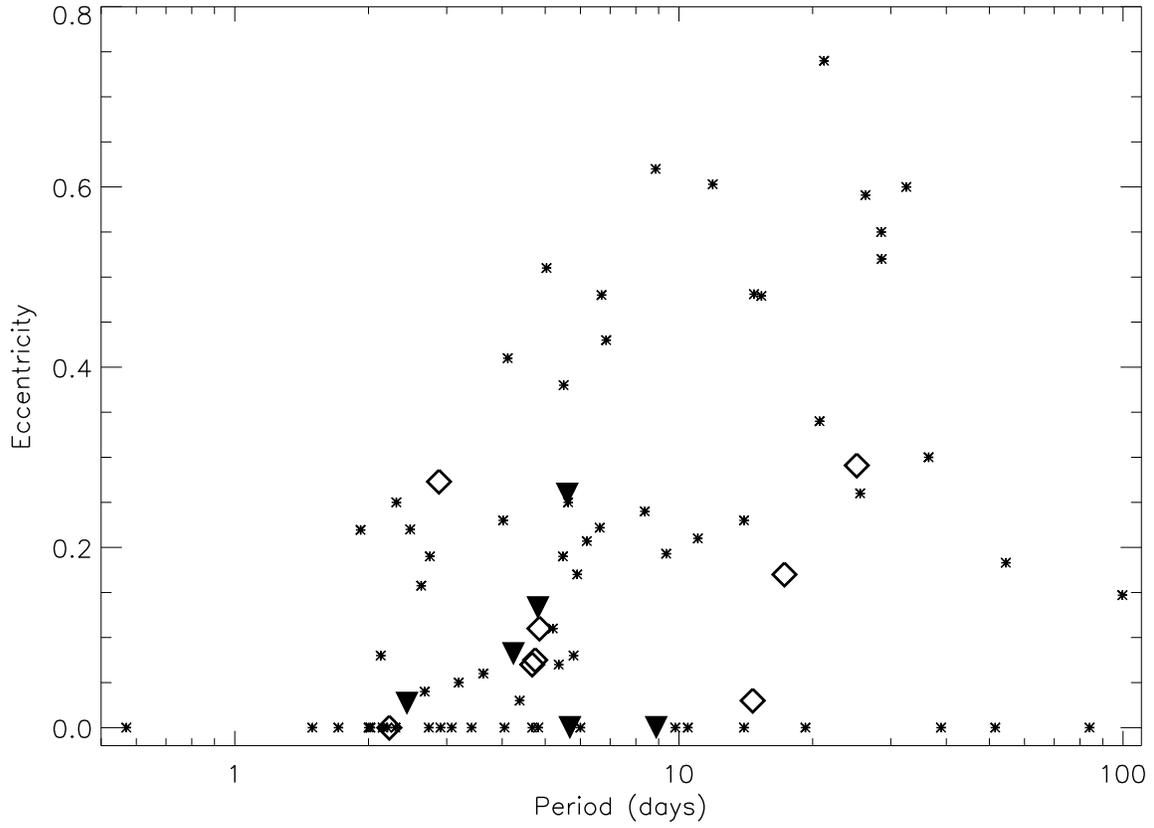}
\caption{Period versus eccentricity for stars in Cyg OB2 (open
diamonds), NGC~6231 (solid triangles) and the OB stars in the 9th
Spectroscopic Binary catalog (asterisks).
\label{pvse}}
\end{figure}


\begin{thebibliography}{}

\bibitem[Bonnell et al.(1998)]{Bonnell98}
Bonnell, I. A., Bate, M.~R., \&\ Zinnecker, H. 1998, MNRAS, 298, 93
\bibitem[Comer\'{o}n et al.(2002)]{Comeron02} 
Comer\'{o}n, F., et al. 2002, A\&A, 389, 874
\bibitem[Comer\'{o}n \&\ Pasquali(2007)]{Comeron07}
Comer\'{o}n, F., \&\ Pasquali, A. 2007, A\&A, 467, 23
\bibitem[Conti \&\ Leep(1974)]{Conti74}
Conti, P. S, Leep, E. M. 1974, ApJ, 193, 113
\bibitem[Contreras et al.(1997)]{Contreras97}
Contreras, M. E., Rodriguez, L. F., Tapia, M., Cardini, D., Emanuele, A., 
Badiali, M., \&\ Persi, P. 1997, ApJ, 488, 153
\bibitem[De Becker et al.(2004)De Becker, Rauw, and Manfroid]{Debeck04}
De Becker, M., Rauw, G., \&\ Manfroid, J. 2004, A\&A, 424, L39
\bibitem[Drilling \& Landolt(2000)]{drilling}
Drilling, J. S., \& Landolt, A. U. 2000, in Astrophysical Quantities, 
ed. A. N. Cox (4th ed.; New York; Springer), 381
\bibitem[Galazutdinov et al.(2000)]{Gala00}
Galazutdinov, G. A., Musaev, F. A., Krelowski, J., \&\ 
Walker, G. A. H. 2000, PASP, 112, 648
\bibitem[Giuricin et al.(1984)]{Giuricin84}
Giuricin, G., Mardirossian, F., \&\ Mezzetti, M. 1984, A\&A, 134, 365
\bibitem[Hanson(2003)]{Hanson03}
Hanson, M. M. 2003, ApJ, 597, 957
\bibitem[Harmanec(2002)]{Harmanec02}
Harmanec, P., Bisikalo, D. V., Boyarchuk, A. A., \&\ Kuznetsov O. A. 2002, A\&A, 396, 937
\bibitem[Herbig(1975)]{Herbig1975}
Herbig, G. H. 1975, ApJ, 196, 129
\bibitem[Herbig \&\ Leka(1991)]{Herbig1991}
Herbig, G. H., \&\ Leka, K. D. 1991, ApJ, 382, 193
\bibitem[Hillwig et al.(2006)]{Hillwig06}
Hillwig, T. C., Gies, D. R., Bagnuolo, W. G., Jr., Huang, W., 
McSwain, M. V., \&\ Wingert, D. W. 2006, ApJ, 639, 1069
\bibitem[Kiminki et al.(2007)]{Kiminki07}
Kiminki, D. C., et al. 2007, ApJ, 664, 1120
\bibitem[Kiminki et al.(2008)]{Kiminki08}
Kiminki, D. C., McSwain, M. V., \&\ Kobulnicky, H. A. 2008, ApJ, 679, 1478
\bibitem[Kinemuchi et al.(2008)]{Karen}
Kinemuchi, K., et al. 2008, in prep 
\bibitem[Kn\"{o}dlseder(2000)]{Kn00}
Kn\"{o}dlseder, J. 2000, A\&A, 360, 539
\bibitem[Kurtz et al.(1991)] {xcsao} 
Kurtz, M. J., Mink, D. J.,
Wyatt, W. F., Fabricant, D. G., Torres, G., Kriss, G. A., \&\ Tonry,
J. L. 1991, in Astronomical Data Analysis Software and Systems I, ASP
Conf. Ser., Vol. 25, eds. D.M. Worrall, C. Biemesderfer, and
J. Barnes, p. 432
\bibitem[Larson(2001)]{Larson01}
Larson, R. B. 2001, IAU Symposium, 200, 93
\bibitem[Lanz \&\ Hubeny(2003)]{LHub2003}
Lanz, T., \&\ Hubeny, I. 2003, ApJS, 146, 417 
\bibitem[Linder et al.(2007)]{Linder07}
Linder, N., Rauw, G., Sana, H., De Becker, M., \&\ Gosset, E. 2007, A\&A, 474, 193
\bibitem[Liu \&\ Yang(2003)]{Liu03}
Liu, Q.-Y.; Yang, Y.-L. 2003, Chinese Journal of Astronomy and Astrophysics, 3, 142
\bibitem[Martins et al.(2005)Martins, Schaerer, and Hillier]{FM05} 
Martins, F., Schaerer, D., \&\ Hillier, D. J. 2005, A\&A, 436, 1049
\bibitem[Massey \&\ Thompson(1991)]{MT91} 
Massey, P., \&\  Thompson, A. B. 1991, AJ, 101, 1408
\bibitem[Massey et al.(2005)]{Massey05}
Massey, P., Puls, J., Pauldrach, A. W. A., Bresolin, F., Kudritzki, 
R. P., \&\ Simon, T. 2005, ApJ, 627, 477
\bibitem[McSwain(2003)]{mcswain03}
McSwain, M. V. 2003, ApJ, 595, 1124
\bibitem[McSwain(2007)]{mcswain07}
McSwain, M. V. 2007, ApJ, 655, 473
\bibitem[Miczaika(1953)]{Mics53}
Miczaika, G. R. 1953, PASP, 65, 141
\bibitem[Milone(1968)]{Milone1968}
Milone, E. F. 1968, AJ, 73, 708
\bibitem[Morbey \& Brosterhus(1974)]{Morbey74}
Morbey, C. L., \&\ Brosterhus, E. B. 1974, PASP, 86, 455
\bibitem[Morton(1991)]{Morton1991}
Morton, D. C. 1991, ApJ, 77, 119
\bibitem[Naz\'{e} et al.(2008)]{Naze08}
Naz\'{e}, Y., De Becker, M., Rauw, G., \&\ Barbieri, C. 2008, A\&A, 483, 543 
\bibitem[Negueruela et al.(2008)]{Neg08}
Negueruela, I., Marco, A., Herrero, A., \&\ Clark, J. S. 2008, A\&A, 487, 575 
\bibitem[\"{O}pik(1924)]{Opik24}
\"{O}pik, E. J. 1924, Tartu Obs. Publ., 25
\bibitem[Otero(2008a)]{NSVSa}
Otero, S. 2008a, Open European Journal on Variable Stars, 83, 1
\bibitem[Otero(2008b)]{NSVSb}
Otero, S. 2008b, Open European Journal on Variable Stars, 91, 1
\bibitem[Penny et al.(2008)]{Penny08}
Penny, L. R., Ouzts, C., \&\ Gies, D. R. 2008, 681, 554
\bibitem[Pourbaix et al.(2004)]{Pourbaix04}
Pourbaix D., Tokovinin A. A., Batten A. H., Fekel F. C., Hartkopf W. I., 
Levato H., Morrell N. I., Torres G., Udry, S. 2004, A\&A, 424, 727
\bibitem[Pigulski \&\ Kolaczkowski(1998)Pigulski \&\ Kolaczkowski]{PJ98}
Pigulski, A., \&\ Kolaczkowski, Z. 1998, MNRAS, 298, 753
\bibitem[Rauw et al.(1999)]{Rauw99}
Rauw, G., Vreux, J. M., \&\ Bohannan, B. 1999, ApJ 517, 416
\bibitem[Rios \&\ DeGioia-Eastwood(2004)]{Rios04}
Rios, L. Y., \&\ DeGioia-Eastwood, K. 2004, BAAS, 205, No. 09.05
\bibitem[Roberts et al.(1987)]{Roberts87}
Roberts, D. H., Leh\'{a}r, J., \& Dreher, J. W. 1987, AJ, 93, 968 
\bibitem[Romano(1969)]{Romano69}
Romano, G. 1969, MmSAI, 40, 375
\bibitem[Salpeter(1955)]{Salpeter55}
Salpeter, E. E. 1955, ApJ, 121, 161
\bibitem[Sana et al.(2008)]{Sana08}
Sana, H., Gosset, E., Naz\'{e} Y., Rauw, G., \&\ Linder, N. 2008, MNRAS, 386, 447
\bibitem[Schulte(1956)]{Schulte56}
Schulte, D. H. 1956, ApJ, 123, 250
\bibitem[Stefanik et al.(1999)]{Stefanik1999}
Stefanik, R. P., Latham, D. W., Torres, G. 1999, ASPC, 185, 354
\bibitem[Stroud et al.(2009)]{Stroud09}
Stroud, V. E., Clark, J.S., Negueruela, I. , Roche, P., 
\&\ Norton, A.J. 2009, A\&A, submitted
\bibitem[Walborn(1973)]{Wal73}
Walborn, N. R. 1973, ApJ, 180, L35
\bibitem[Walborn \&\ Fitzpatrick(1990)]{WF90}
Walborn, N. R., \&\ Fitzpatrick, E. L. 1990, PASP, 102, 379
\bibitem[Walborn \&\ Howarth(2000)]{Walborn00}
Walborn, N. R., \&\ Howarth, I. D. 2000, PASP, 112, 1446
\bibitem[Williams et al.(2008)]{Williams08}
Williams, et al. 2008, ApJ, 682, 492
\bibitem[Wilson(1948)]{Wilson48}
Wilson, O. C. 1948, PASP, 60, 385
\bibitem[Wilson \&\ Abt(1951)]{Wilson51}
Wilson, O. C., \&\ Abt, A. 1951, ApJ, 144, 477
\bibitem[Wozniak et al.(2004)]{Wozniak04}
Wozniak, P. R., et al. 2004, AJ, 127, 2436, 
Northern Sky Variability Survey: Public Data Release 
\end{thebibliography}
\end{document}